\newif\ifShowKeys
\ShowKeystrue
\documentclass[11pt,a4paper]{article} 
\usepackage[height=8.8in,width=6.45in]{geometry}

\usepackage[hidelinks]{hyperref}

\usepackage{tocloft}
\usepackage{amsmath, amssymb,amsthm}
\usepackage{amsthm}
\numberwithin{equation}{section}

\usepackage{bm,environ,mathrsfs,array,arydshln}
\usepackage{booktabs,float,slashed,hyperref}
\usepackage[mathcal]{euscript}
\usepackage{tensor} 						% Ratcliffe package to write tensors
\usepackage{mathabx}
\usepackage{simpler-wick}

\usepackage{aurical}
\usepackage[T1]{fontenc}

\usepackage[nodayofweek]{date time}
\usepackage{graphicx,epsfig,epic}
\usepackage{tikz} 
\usetikzlibrary{decorations.pathreplacing,decorations.markings,snakes}
\tikzset{middlearrow/.style={decoration={markings, mark= at position 0.5 with {\arrow{#1}} ,
}, postaction={decorate}}}

\usepackage{framed}						% for shaded equations \begin{shaded}...\end{shaded}
\definecolor{shadecolor}{rgb}{0.95,0.95,0.97}

\usepackage{comment}

\allowdisplaybreaks

\newcommand{\bs}{\begin{shaded}}
\newcommand{\es}{\end{shaded}}
\def\ba#1\ea{\begin{align}#1\end{align}}		% very clever way to bypass the known problem...
\newcommand{\be}{\begin{equation}}
\newcommand{\ee}{\end{equation}}
\newcommand{\mc}{\mathcal }

\newcommand{\la}{\label}

\newcommand{\lp}{\notag \\ & }

\DeclareMathOperator{\tr}{\text{tr}}

\newcommand{\cf}{\textit{cf.} }
\newcommand{\ie}{\textit{i.e.} }

\makeatletter
\DeclareFontFamily{OMX}{MnSymbolE}{}
\DeclareSymbolFont{MnLargeSymbols}{OMX}{MnSymbolE}{m}{n}
\SetSymbolFont{MnLargeSymbols}{bold}{OMX}{MnSymbolE}{b}{n}
\DeclareFontShape{OMX}{MnSymbolE}{m}{n}{
<-6>  MnSymbolE5
   <6-7>  MnSymbolE6
   <7-8>  MnSymbolE7
   <8-9>  MnSymbolE8
   <9-10> MnSymbolE9
  <10-12> MnSymbolE10
  <12->   MnSymbolE12
}{}
\DeclareFontShape{OMX}{MnSymbolE}{b}{n}{
<-6>  MnSymbolE-Bold5
   <6-7>  MnSymbolE-Bold6
   <7-8>  MnSymbolE-Bold7
   <8-9>  MnSymbolE-Bold8
   <9-10> MnSymbolE-Bold9
  <10-12> MnSymbolE-Bold10
  <12->   MnSymbolE-Bold12
}{}

\let\llangle\@undefined
\let\rrangle\@undefined
\DeclareMathDelimiter{\llangle}{\mathopen}%
 {MnLargeSymbols}{'164}{MnLargeSymbols}{'164}
\DeclareMathDelimiter{\rrangle}{\mathclose}%
 {MnLargeSymbols}{'171}{MnLargeSymbols}{'171}
\makeatother

\newmuskip\pFqmuskip

\newcommand*\pFq[6][8]{%
  \begingroup % only local assignments
  \pFqmuskip=#1mu\relax
  \mathchardef\normalcomma=\mathcode`,
  \mathcode`\,=\string"8000
  \begingroup\lccode`\~=`\,
  \lowercase{\endgroup\let~}\pFqcomma
  {}_{#2}F_{#3}{\left[\genfrac..{0pt}{}{#4}{#5};#6\right]}%
  \endgroup
}
\newcommand{\pFqcomma}{{\normalcomma}\mskip\pFqmuskip}

\def\XXint#1#2#3{{\setbox0=\hbox{$#1{#2#3}{\int}$}
     \vcenter{\hbox{$#2#3$}}\kern-.5\wd0}}

\def \foot{\footnote}\def \ci{cite}\def \l {\lambda}\def \iffa {\iffalse}
\def \RR {{R}} \def \ov {\over }\def \a  {\alpha} \def \ha {{1\ov 2}}
\def \ed {\end{document}}

\def \l {\lambda}
\def\foot{\footnote}
\def \adss {${\rm AdS}_5 \times S^5~$ }
\def \ov {\over}
\def \tr {{\rm tr\,}}
\def \ha {{1 \over 2}}
\def \td {\tilde}
\def \ci{\cite}

\def \N  {{\cal N}}

\def \te {\textstyle} 

\def \RR {{L}}

\newcommand{\rf}[1]{(\ref{#1})}

\newcommand{\sql}{\sqrt\lambda}
\newcommand{\gs}{g_{\text{s}}}

\newcommand{\vev}[1]{\left\langle  #1 \right\rangle}

\newcommand{\sym}{\mc N=4}
\newcommand{\orientif}{\rm orient}

\begin{document}

\begin{titlepage}

%\date{\currenttime}
%\begin{flushright}\boxed{\small{\tt \today \ \ - \ \  \currenttime }}\end{flushright}

\begin{tabbing}
\hspace*{11.5cm} \=  \kill % set the tabbings
\> Imperial-TP-AT-2021-02 %additional information:  \\
%\> none
\end{tabbing}

\vspace*{3mm}
\begin{center}
{\Large\sc  BPS Wilson loop in $\mc N=2$ superconformal}\vskip 7pt
{\Large\sc   $SU(N)$ ``orientifold''  gauge  theory}\vskip 7pt
{\Large\sc  and weak-strong coupling interpolation}\vskip 7pt

\vspace*{4mm}

{\Large  M. Beccaria${}^{\,a}$, G.V. Dunne${}^{\,b}$ and   A.A. Tseytlin${}^{\,c,}$\footnote{\ Also at the Institute for Theoretical and Mathematical Physics, MSU and Lebedev Institute, Moscow}
}

\vspace*{5mm}
	
${}^a$ Universit\`a del Salento, Dipartimento di Matematica e Fisica \textit{Ennio De Giorgi},\\ 
		and I.N.F.N. - sezione di Lecce, Via Arnesano, I-73100 Lecce, Italy
			\vskip 0.15cm
${}^b$ Department of Physics, University of Connecticut, Storrs, CT 06269-3046, USA
			\vskip 0.15cm
${}^c$ Blackett Laboratory, Imperial College London SW7 2AZ, U.K.
			\vskip 0.15cm
			
\vskip 0.2cm
	{\small
		E-mail:
		\texttt{matteo.beccaria@le.infn.it,  gerald.dunne@uconn.edu,    tseytlin@imperial.ac.uk}
	}
\vspace*{0.5cm}
\end{center}

\begin{abstract}  
We consider the  expectation value  $\langle\mc  W \rangle $  of the circular  BPS Wilson loop  in  $\N=2$ superconformal  $SU(N)$ gauge theory 
containing a vector multiplet coupled to two  hypermultiplets  in rank-2 symmetric and  antisymmetric representations. 
This theory admits a regular    large $N$ expansion, 
 is planar-equivalent to $\N=4$ SYM   theory and is expected to be dual  to a certain orbifold/orientifold projection of 
AdS$_5\times S^5$   superstring theory. 
On the string theory  side   $\langle\mc  W \rangle $ is represented by the  path integral expanded near  the same AdS$_2$ minimal 
surface as in the maximally supersymmetric  case.  Following  the string  theory argument  in  arXiv:2007.08512, we  
suggest     that as  in the $\N=4$ SYM   case and in   the 
$\N=2$  $SU(N) \times SU(N)$  superconformal  quiver theory  discussed in arXiv:2102.07696, 
the  coefficient  of the  leading  non-planar  $1/N^2$ correction in $\langle\mc  W \rangle $  should  have the universal   $\l^{3/2}$
scaling at large  't Hooft coupling. 
We confirm this prediction  by starting with the localization matrix model  representation for 
 $\langle\mc  W \rangle $. %  and analytically derive  the large $\lambda$ limit of the  coefficient of the $1/N^2$  term. 
We   complement  the analytic  derivation  of    the $\l^{3/2}$ scaling by  a numerical  high-precision resummation and extrapolation of  the weak-coupling expansion using  conformal mapping improved Padé analysis.
\end{abstract}
\vskip 0.5cm
	{
		%Keywords: {\sc insert here keywords}
	}
\end{titlepage}

\iffalse
%%%%%%%%%FOR SUBMISSION  %%%%%%%%%%%%%%%%%%%%%%%%%%%%%%%%%%

We consider the  expectation value  $\langle \cal  W \rangle$  of the  circular  BPS Wilson loop
  in  ${\cal N}=2$ superconformal  $SU(N)$ gauge theory 
containing a vector multiplet coupled to two  hypermultiplets  in rank-2 symmetric and  antisymmetric representations. 
This model  admits a regular    large $N$ expansion, 
 is planar-equivalent to ${\cal N}=4$ SYM   theory and is expected to be dual  to a certain orbifold/orientifold projection of 
AdS$_5\times S^5$   superstring theory.  On the string  theory side   $\langle\cal  W \rangle $ is represented by the  path integral expanded near  the same AdS$_2$ minimal  surface as in the maximally supersymmetric  case.  Following  the string  theory argument  in  arXiv:2007.08512, we   suggest     that as  in the ${\cal N}=4$ SYM   case and in   the 
${\cal N}=2$  $SU(N) \times SU(N)$  superconformal  quiver theory  discussed in arXiv:2102.07696, 
the  coefficient  of the  leading  non-planar  $1/N^2$ correction in $\langle\cal  W \rangle $  should  have the universal   $\lambda^{3/2}$
scaling at large  't Hooft coupling.  We confirm this prediction  by starting with the localization matrix model  representation for 
 $\langle\cal  W \rangle $. We   complement  the analytic  derivation  of   the $\lambda^{3/2}$ scaling by  a numerical  high-precision resummation and extrapolation of  the weak-coupling expansion using  conformal mapping improved Padé analysis.

%%%%%%%%%%%%%%%%%%%%%%
\fi

\tableofcontents
\vspace{1cm}

\def \no { \nonumber}
\def \S {{\rm S}} \def \ss{{\rm s}} \def \jj {{\rm j}}

\setcounter{footnote}{0}

\section{Introduction}

Wilson loops are an important class of observables in gauge and string theory that, in particular,  
 help clarifying the interpolation between the weak and strong coupling regimes from the AdS/CFT perspective.
In several supersymmetric gauge theories it is possible to compute expectation values  of 
BPS Wilson loops  using 
localization  in terms of  matrix model integrals %A whose small $\lambda$ expansion can be developed perturbatively
(see, e.g.,  \cite{Pestun:2016zxk}). 

In the maximally  supersymmetric $\mc N=4$    $SU(N)$ gauge theory 
the corresponding  matrix model can be solved for any  gauge coupling and gauge group rank
%as in the well known $SU(N)$ gauge group  case 
\cite{Drukker:2000rr}.\foot{The same is true also for the $\mc N=4$  theory with    $SO(N)$  and $USp(N)$ gauge groups 
 \ci{Fiol:2014fla,Giombi:2020kvo}.}
 This allows, in particular,  to study  the  strong coupling  limit  of the  coefficients in the $1/N$ expansion,  
 providing  a  possibility to compare to the large tension limit  of the  coefficients in the expansion in   powers of string coupling (genus) 
  on the dual \adss string theory side, and  thus leading to highly non-trivial checks of AdS/CFT duality \cite{Giombi:2020mhz,Beccaria:2020ykg,Beccaria:2021alk}.

Localization method   applies also 
to  a large class of  gauge theories  with reduced $\mc N=2$ supersymmetry, 
 but the associated matrix models have 
non-polynomial potentials and are not  directly solvable. 
%A
While developing the  small $\l$ expansion  is straightforward, 
   extracting the strong coupling limit of the gauge theory observables is a   non-trivial problem. 
At leading order in  the large $N$  expansion 
this requires a Wiener-Hopf analysis of the matrix model as  first exploited in  \cite{Passerini:2011fe} for $\mc N=2$ $SU(N)$ SYM 
with $N_{F}=2N$ fundamental hypermultiplets, and later generalized to other superconformal Lagrangian models admitting a large $N$ limit 
\cite{Russo:2013kea,Zarembo:2014ooa,Baggio:2014sna,Baggio:2015vxa,Baggio:2016skg,Fiol:2015mrp,Zarembo:2016bbk}. 
The  methods used at leading planar level  are not,  however,  applicable to the analysis of the higher 
$1/N$ corrections. 

An interesting class of models where $1/N$  corrections  happen to 
 be more tractable is  that of $\mc N=2$ superconformal  gauge theories  with $SU(N)$  gauge group 
whose leading   large $N$ limit is equivalent  (in a particular ``common''  sector) to that of the $\mc N=4$    SYM   theory. 
One such example is the $SU(N)\times SU(N)$ quiver gauge theory 
with  bi-fundamental  hypermultiplets and equal gauge couplings. 
This $\mc N=2$ theory  may be interpreted as  a  $\mathbb Z_{2}$ orbifold of  the $\mc N=4$  $SU(2N)$ SYM and is 
dual to  superstring theory on AdS$_{5}\times (S^{5}/\mathbb Z_{2})$ \cite{Kachru:1998ys}.

Let us first review  some basic results  about the expectation value $\vev{\mc W} $ of 
 circular $\frac{1}{2}$-BPS Wilson loop  in  $\mc N=4$ SYM theory. 
%Introducing the 't Hooft coupling $\lambda = g^{2}_{\rm YM}\,N$, 
%in $SU(N)$ $\mc N=4$ SYM, 
Its  planar limit  is given by   %of the $\frac{1}{2}$-BPS circular Wilson loop  
is \cite{Erickson:2000af,Drukker:2000rr,Pestun:2007rz}
\ba
\la{1.1}
\vev{\mc W}_0 = 
\frac{2\,N}{\sqrt\lambda}\,I_{1}(\sqrt\lambda) = \sqrt\frac{2}{\pi}\,N\,\lambda^{-3/4}\,e^{\sqrt\lambda}
\Big[1+\mc O\Big(\frac{1}{\sqrt\lambda}\Big)\Big]  \ , \qquad \qquad   \lambda = g^{2}_{_{\rm YM}}\,N \ .   %+\cdots \ . 
\ea
The relative weight of the leading $1/N$ correction to $\vev{\mc W}$ with respect to the planar result may be 
represented as  %written in terms of the 
%function $q(\lambda)$ defined by the expansion
\be
\la{1.2}
\frac{\vev{\mc W}}{\vev{\mc W}_0} = 1+\frac{1}{N^{2}}\,q(\lambda)+\mc O\Big(\frac{1}{N^{4}}\Big) \ ,  
\ee
%The where  the function  $q(\lambda)$ 
where the function $q(\lambda)$  (and its  analogs  in the $\mc N=2$  models  with planar equivalence to $\mc N=4$  SYM) 
will be our main interest below. Starting   with the general (Laguerre polynomial)  expression  for   $\vev{\mc W} $   in  $\mc N=4$   $SU(N)$ SYM  \cite{Drukker:2000rr}
one finds  for the  leading terms in $q(\lambda)$ at  weak and strong coupling\foot{In what follows the label ``$\sym$''  will always refer to the   $\mc N=4$   $SU(N)$ SYM   expression.  }  
\be
\la{1.3}
\qquad q^{\sym}(\lambda) = \frac{\lambda}{96}\Big[\frac{\sqrt\lambda\,I_{2}(\sqrt\lambda)}{I_{1}(\sqrt\lambda)}-12\Big] = \begin{cases}
-\frac{1}{8}\lambda+\frac{1}{384}\lambda^{2} -\frac{1}{9216}\,\lambda^{3}+\mc O(\lambda^{4}), &\qquad  \lambda \to 0, \\
\frac{1}{96}\lambda^{3/2}-\frac{9}{64}\lambda+\frac{1}{256}\lambda^{1/2}+\mc O(1), &\qquad  \lambda \to \infty \ . 
\end{cases}
\ee
The $\lambda^{3/2}$  scaling of  $q^{\sym}$   at $\lambda\gg 1$ has a  string interpretation. The 
 string coupling and tension for the dual string  theory on AdS$_{5}\times S^{5}$ are defined as 
\be
\la{1.4}
\gs =  \frac{\lambda}{4\pi N},\qquad \qquad T = \frac{\RR^2}{2\pi \a'}=\frac{\sqrt\lambda}{2\pi}.
\ee
As was argued in \cite{Giombi:2020mhz},     the   leading  large $T$ dependence  of the  string theory 
expectation value for $\vev{\mc W} $  at each order in $\gs$ is  also    
  controlled by the Euler number, $\chi=1-2p$,   (or genus $p$) 
 of the string world sheet, \ie 
\be\la{1.5}
\vev{\mc W} =\sum_{p=0}^{\infty} \vev{\mc W}_p =  e^{2\pi\,T}\sum_{p=0}^{\infty}c_{p}\,\Big(\frac{\gs}{\sqrt T}\Big)^{2p-1}\Big[1+\mc O\Big(T^{-1}\Big)\Big] \ .
\ee
Written in terms of $N$ and $\l$ in  \rf{1.4}    this reads  ($c_p'= { c_p \ov ( 8 \pi)^{p-1/2}}$)
  \be
\la{1.6}
\vev{\mc W}^{} =  N\,e^{\sqrt\lambda}\sum_{p=0}^{\infty}
 c'_p \frac{\lambda^{\frac{6p-3}{4}}}{N^{2p}}\Big[1+\mc O\Big(\frac{1}{\sqrt\lambda}\Big)\Big] \ , 
\ee
and thus 
 matches the  structure of the  $1/N$  expansion of the  exact    $\mc N=4$ SYM  result   \cite{Drukker:2000rr}.  
 In particular, comparing to \rf{1.2}, 
 we have,  in agreement with (\ref{1.3}), 
\be
\la{1.7}
\frac{1}{N^{2}}\,q^{\sym}(\lambda)\  \stackrel{\lambda\gg 1}{\sim}\ \frac{\gs^{2}}{T}\,\ \propto\,\  \frac{\lambda^{3/2}}{N^{2}} \ . 
\ee
The  discussion in   \cite{Giombi:2020mhz} leading to  \rf{1.5}  
% universality of the  structure of the expansion in \rf{1.5}:
  relied only on  the fact that one expands near  the AdS$_2$ minimal  surface 
embedded into the AdS$_3$ part of AdS$_5$ space;   thus  it should apply    not only  to \adss   superstring but also 
to its  closely related  orbifold   and orientifold  modifications 
  based on AdS$_5 \times S'^5$   where $S'^5$  is  locally a 5-sphere. 
  Indeed, since the   fluctuations of string world sheet fields 
   related to $S'^5$    remain ``massless'', the 
 reasoning    \cite{Giombi:2020mhz}   determining   the tension dependence from  the 
 way how  the AdS$_5$  radius  appears in the  1-loop (leading large $T$) string partition function  should not change. 
 
  In \cite{Beccaria:2021ksw} it was  argued  that this  should   apply, in particular, 
   to the orbifold   ${\rm AdS}_5\times (S^{5}/\mathbb Z_{2})$   theory and evidence for the validity 
   of \rf{1.5},\rf{1.6}    was provided at the first non-trivial $1/N^2$ order. 
To recall, 
in the $SU(N)\times SU(N)$ orbifold model, 
%it is possible to define a $\frac{1}{2}$-BPS circular Wilson loop whose planar 
%expectation value is equal to the $\mc N=4$ SYM expression.
for each of the two $SU(N)$  factors, it is possible to define the  $\frac{1}{2}$-BPS  circular Wilson loops 
coupled to  the associated gauge and scalar fields and $\vev{\mc W_{1}} = \vev{\mc W_{2}}\equiv\vev{\mc W}^{\rm orb}$. 
At the leading planar  level  one has 
$\vev{\mc W}^{\rm orb}_{N\to\infty} =  \vev{\mc W}^{\sym}_{N\to\infty} =  \vev{\mc W}_0$
 \cite{Rey:2010ry,Zarembo:2020tpf}.\footnote{The strong coupling limit of the planar  expectation value 
 of a similar 
  Wilson  loop in  quiver   gauge theory  with  unequal 
 gauge couplings  was  solved  in \cite{Zarembo:2020tpf}, see also \cite{Mitev:2014yba,Mitev:2015oty,Ouyang:2020hwd}.}
%At intermediate coupling, the
Starting from  the  corresponding localization matrix model representation for   $\vev{\mc W}^{\rm orb}$, 
a numerical analysis of the $q^{\rm orb}(\lambda)$ function, defined as in (\ref{1.3}), 
extrapolated to large $\l$  values 
 gave   the  following estimate \cite{Beccaria:2021ksw}  
%\be\la{1.1}
%\mc W_\aa = \tr\, \mc P\,\exp \Big[
%\oint ds\,(i\,\dot x^{\mu}\,A_{\mu\,\aa}+|\dot x|\,\Phi_\aa) \Big] \ , \qquad \langle{\mc W_1} \rangle= \langle{\mc W_2}\rangle\equiv \vev{\mc W}^{\rm orb}.
%\ee
%, 
\ba\la{18}
q^{\rm orb}(\lambda)   \ \stackrel{\lambda\gg 1}{=} C \,  \lambda^{\eta}, \qquad  
\eta = 1.49(2), \ \qquad  C\simeq - 0.0049(5)\ .
\ea
The value of the  asymptotic exponent $\eta$  is thus quite consistent with  the string theory expectation 3/2  in  (\ref{1.7}).

\

%%%%%%%%%%%%%%%%%%%%%%%%%%%%%%%%%%%%%%%%%%
%AA

In this paper  we shall consider   another  $\mc N=2$ superconformal  model where  the  structure of the  large $N$, strong coupling expansion 
 should be  of the same universal form as in \rf{1.5}.  We shall  confirm this %physical 
 expectation with
 % shall be able to  check this  by presenting not only 
 an analytic  argument  for the  strong-coupling scaling in \rf{1.7},  in addition to 
 numerical evidence based on high-precision extrapolation from the weak to strong coupling regimes.
 
 This theory 
is the  $SU(N)$ gauge theory with $\mc N=2$ vector multiplet  coupled to  two hypermultiplets -- 
in  rank-2 symmetric  and antisymmetric  $SU(N)$ representations.\footnote{It 
 is one of the  five cases of   4d  $\mc N=2$ superconformal theories  with gauge group $SU(N)$ 
 defined  for an arbitrary value of  $N$   \cite{Howe:1983wj,Koh:1983ir}. 
% and thus admitting the $1/N$ expansion.
} 
This model admits a regular 't Hooft large $N$ expansion and  %at the planar level  is 
 its  string theory  dual 
is expected to be  a particular orientifold of AdS$_{5}\times S^{5} $
  type IIB   superstring theory \cite{Park:1999ep,Ennes:2000fu}.
For that   reason  in what follows we shall   refer to this $\mc N=2$  gauge theory  as  the  ``orientifold theory''.

%and is dual to IIB superstring on a specific  orientifold 
%AdS$_{5}\times S^{5}/(\mathbb Z_{2}^{\rm orient}\times \mathbb Z_{2}^{\rm orb})$  \cite{Park:1998zh,Ennes:2000fu}. 
%In the following we shall refer to this theory  as ``orientifold theory''.
To recall, in  $\mc N=2$ gauge theories the  $\beta$-function  for the gauge coupling has only the 1-loop  contribution: 
%perturbative corrections.
in  a model  with $N_{F}$ hypermultiplets in the fundamental, $N_{S}$ in  rank-2 symmetric, and $N_{A}$  in  rank-2 antisymmetric  representations one has 
$\beta_{\rm 1-loop} = 2N-N_{F}-N_{S}(N+2)-N_{A}(N-2)$. It   thus  vanishes  for the 
 orientifold theory   where $N_F=0, \ N_A=N_S=1$. 
Let us also mention for completeness  that  the 4d  Weyl anomaly coefficients a  and  c 
for an $\mc N=2$ theory with $n_{V}$ vector and $n_{H}$ hyper multiplets
 are  given by ${\rm a} = \tfrac{5}{24}\,n_{V}+\tfrac{1}{24}\,n_{H}$,\  ${\rm c} = \tfrac{1}{6}\,n_{V}+\tfrac{1}{12}\,n_{H}$,
 %, and in particular ${\rm c}-{\rm a} = \tfrac{1}{24}(n_{H}-n_{V})$.
so that in the present case   with 
$n_{V}=N^{2}-1$ and $n_{H} = \frac{1}{2}N(N+1)+\frac{1}{2}N(N-1) = N^{2}$, we get 
${\rm a} = \tfrac{1}{4}N^{2}-\tfrac{5}{24}$, \ ${\rm c} = \tfrac{1}{4}N^{2}-\tfrac{1}{6}$.
%, and ${\rm c}-{\rm a}=\tfrac{1}{24}$.
Thus a and c   are equal at the leading    $N^{2}$ order  which is  consistent % a necessary condition for the 
with the existence of a  well  defined holographic dual.\foot{Let us also   note that it  should be possible 
to reproduce the  subleading  terms   in  a  and c  on the dual  orientifold string theory side   by summing up 
the 1-loop contributions of the ``massless'' $D=10$ supergravity  fields 
 (corresponding to short multiplets  represented by  towers of  Kaluza-Klein  modes)   similarly to how that was done in the 
 case of the    $\mc N=4$ $SU(N)$  SYM theory  (where  a=c$={1\over 4} N^2- {1\over 4}$) \cite{Beccaria:2014xda}  and 
 some  orbifold theories  \cite{Ardehali:2013xya}.}
 %of the anomaly coefficients should match the corresponding total quantity from the short multiplets appearing in 
%the $AdS_{5}$ Kaluza-Klein towers of the orientifold string theory. 
%It would be interesting to check holographically these shifts as was done in 
%for $\mc N=4$ SYM, see also \cite{Ardehali:2013xya}.

 Our aim will be   to consider  the 
   expectation value of the $\frac{1}{2}$-BPS circular Wilson loop in the orientifold theory.  
   At the leading large $N$ order it is  the same as in the $\mc N=4$  SYM  theory in \rf{1.1}.\footnote{This is a 
   manifestation of the planar equivalence  between  the orientifold theory  and  the  $\mc N=4$  SYM  in  the ``untwisted''  sector.
   For a detailed discussion of planar equivalence violations in ``odd'' sectors  see \cite{Beccaria:2020hgy}.}
   The main focus  will be  on the 
    leading non-planar correction    represented by the  function $q^{\orientif}(\lambda)$ defined  as in (\ref{1.2}).
%different than those of $\mc N=4$ SYM or of the orbifold.

The string   dual  of   this  $\mc N=2$ model 
%On the string side, the orientifold theory is dual to a  special 
is the   type IIB   superstring  theory  defined   on  the orientifold  
 AdS$_{5}\times S^{5}/G_{\rm orient} $    \cite{Ennes:2000fu}.
%  to the 
%  $\mc N=2$   superconformal   gauge  theory involving   $SU(N)$  $\N=2$ vector multiplet  coupled to 
%two  hypermultiplets --  in  rank-2 symmetric  and antisymmetric  $SU(N)$ representations. 
%This theory   admits a regular 't Hooft large $N$  limit   and thus  is similar to the quiver  theory discussed   above.  It 
 %   should be dual to the type IIB superstring  on a particular  orientifold 
%AdS$_{5}\times S^{5}/(\mathbb Z_{2}^{\rm orient}\times \mathbb Z_{2}^{\rm orb})$
% (see \cite{Ennes:2000fu}). 
   Here $G_{\rm orient} =\mathbb Z_{2}^{\rm orb} \times  \mathbb Z_{2}^{\rm orient}$,  
   where $Z_{2}^{\rm orient}$  in addition to  the  target space coordinate 
    inversions (in directions transverse to the original D3-branes) 
    involves the product  
    world-sheet parity operator $\Omega$  and $(-1)^{F_L}$. 
    The compact part of the 10d space  $S'^5= S^{5}/G_{\rm orient}$ is  different from $S^5$   only by special  identifications of the angular coordinates
   \cite{Ennes:2000fu}:
   
   \noindent
    \  $ds'^2_5= d \theta_1^2 +\sin^2\theta_1\,  d \phi_3^2 +  \cos^2\theta_1\, ds'^2_3 , \ \ \ \ 
   ds'^2_3 = d \theta_2^2 +\sin^2\theta_2\,  d \phi_2^2 +  \cos^2\theta_2\, d\phi_1^2$, \  \
   
   \noindent
   $\theta_{1}\equiv \theta_1 + {\pi \ov 2}, \ \theta_{2}\equiv \theta_2 + {\pi \ov 2},\ \
   \phi_1 \equiv \phi_1 + {\pi \ov 2},\ \   \phi_2 \equiv \phi_2 - {\pi \ov 2},\ \
    \phi_3 \equiv \phi_3 + {\pi}$.

The dual string theory description of the circular  Wilson loop is  based again on the string partition   function 
 expanded near  the AdS$_2$   minimal surface  embedded in AdS$_5$.  %As  already mentioned   above, 
    As the  UV  divergent part of the 1-loop  fluctuation determinants   \cite{Drukker:2000ep} near this   minimal surface
     should   not be   sensitive to the global identifications in the $S'^5$ part of the orientifold  geometry,    the argument in 
     \cite{Giombi:2020mhz}   leading to the   universal structure of   strong-coupling expansion 
     \rf{1.5},\rf{1.6}  should apply not only to the   original \adss or orbifold  theory considered  in  \cite{Beccaria:2021ksw} 
      but also to this   orientifold theory as well.

     Below   we shall provide evidence for  this, i.e.   for the validity of \rf{1.5},\rf{1.6}, 
       on  the dual orientifold gauge theory side 
     by showing that   the localization matrix  model representation for the  circular BPS Wilson loop 
     implies   that the $1/N^2$ term in \rf{1.2}  indeed 
     scales as  in  (\ref{1.7})   at the  leading order  at strong coupling, \ie
 \be
 \la{1.9}
 q^{\orientif}(\lambda) \stackrel{\lambda\gg 1}{\sim} \lambda^{3/2} \ .
 \ee
%which   is  in agreement with
%%%%%%%%%%%%%%%%%%%%%%%%%%%%%%%%%%%%%
%\subsection*{Summary of the results}

\

%A
Let us  briefly  summarize our main results. The aim will be  to present a detailed study of 
the $1/N^2$ coefficient  in (\ref{1.2}) in the orientifold theory, i.e.  of $q^{\orientif}(\lambda)$.
%proving   (\ref{1.9})   and checking  this also by a numerical evaluation. 
As in the orbifold theory discussed  in \cite{Beccaria:2021ksw}, from the 
matrix model representation  for  the  Wilson loop  in the  orientifold   theory  one can relate the difference 
between  $ q^{\orientif}(\lambda) $   and $ q^{\sym}(\lambda)$
%$\Delta q(\lambda)$ defined by 
\be\la{1.10}
\Delta q(\lambda) \equiv q^{\orientif}(\lambda) - q^{\sym}(\lambda)  \ , 
\ee
to the  $N \to \infty$  limit of the   difference  of the corresponding free energies\foot{Note that   while the  individual free energies  on $S^4$ are, in general, scheme-dependent, their  difference $\Delta F$ is scheme-independent.
Earlier discussion of leading terms in perturbative expansion 
in Wilson loop and free energy in this theory was in \cite{Fiol:2020bhf}.
} 
\ba
\la{1.11}
\Delta q(\lambda) &= - \frac{\lambda^{2}}{4}\frac{d}{d\lambda}\,\Delta F(\lambda)\ , 
\qquad \qquad \Delta F(\lambda) \equiv    \lim_{N\to \infty }   \Delta F(\lambda; N )\ ,  \\
\la{1.12}
\Delta F(\lambda; N ) &\equiv  F^{\orientif}(\lambda; N)-F^{\sym}(\lambda; N) = -\log\frac{Z^{\orientif}(\lambda; N)}{Z^{\sym}(\lambda; N)}  \ . 
\ea
Here $Z^{\orientif}$  and  $Z^{\sym}$ are the corresponding partition functions on $S^{4}$.  
The function  $\Delta F(\l)$ turns out to have the following  weak coupling expansion 
\ba
 \Delta F(\lambda) = 
&5 \zeta _5 \hat{\lambda }^3-\tfrac{105}{2} \zeta _7 \hat{\lambda 
}^4+441 \zeta _9 \hat{\lambda }^5 -(25 \zeta _5^2+3465 \zeta _{11}) \hat{\lambda }^6    \nonumber  \\  
&\ \ 
+\big(525 \zeta _5 \zeta _7+\tfrac{3.6355}{8} \zeta _{13}\big) 
\hat{\lambda }^7 
%+\Big(-\frac{22785 \zeta _7^2}{8}-\frac{8505 \zeta _5 
%\zeta _9}{2}-\frac{6441435 \zeta _{15}}{32}\Big) \hat{\lambda 
%}^8  \lp
%+\Big(\frac{500 \zeta _5^3}{3}+\frac{94815 \zeta _7 \zeta_9}{2}+\frac{63525 \zeta _5 \zeta _{11}}{2} 
%+\frac{12167155 \zeta _{17}}{8}\Big) \hat{\lambda }^9
+\cdots, \qquad \qquad  \hat\lambda=\frac{\lambda}{8\pi^{2}} \ .  \la{1.13}
\ea
%Here we have defined a conveniently scaled coupling parameter
%\ba
%\hat\lambda=\frac{\lambda}{8\pi^{2}}
%\la{lhat}
%\ea
Extracting the strong coupling expansion is much harder.  Since  in the  $\mc N=4$ SYM theory  the matrix   model representation 
implies   that  $F^{\sym}= - \frac{1}{2}\,(N^2-1) \log { \l}  $  \cite{Russo:2012ay}
%v3
(we ignore $\l$-independent  but $N$-dependent constant, cf. \ci{Bourget:2018obm}), 
combining \rf{1.11},\rf{1.12} and \rf{1.9}  we get, as in the orbifold theory case   \cite{Beccaria:2021ksw}, 
      the following prediction  for $F^{\orientif}$ 
\be
\la{1.14}
F^{\orientif}(\lambda; N) \stackrel{\lambda\gg 1}{=} - \frac{1}{2}\,N^2 \log { \l}   +  \big[ c_1\,   \sqrt  \lambda +  \mc O(\log \l)   \big] 
+ \mc O \Big({1\ov N^2}\Big)    \ . 
\ee
The    leading $\mc O(N^2)$   term in (\ref{1.14})  is  implied by  the  planar equivalence  to 
  the $SU(N)$   SYM   theory and should follow     from the leading type IIB supergravity  term  evaluated on 
  ${\rm AdS}_5\times (S^{5}/G_{\rm orient})$.\foot{Planar equivalence    also   implies  that   like in the $\mc N=4$ SYM theory 
 this  leading $N^2$ term should not get    string  $\frac{1}{\sql}$ corrections: they should still vanish on
 ${\rm AdS}_5\times (S^{5}/G_{\orientif})$.}
 %v3
 Let us   note that  as  one  can show  from the localization  matrix  model, 
  large $N$  expansion of free energy $F^{\orientif}(\lambda; N) $
 and of the corresponding Wilson loop will contain only even powers of $1/N$.
 This may be somewhat  surprising from the dual string theory 
 point of view  where  in the orientifold case one   should  in general get cross-cup contributions  with  odd powers of $g_s$. 
 
%In the orbifold theory, we  obtained the coefficient $c_{1}$ in (\ref{1.14}) by a numerical Monte Carlo simulation.
%In the orientifold theory, the  matrix model is more tractable and we

%v3
Below  we  will analytically  derive  the $ \sqrt  \lambda $  term in \rf{1.14} and thus in $\Delta F$ finding  that  
%presence of the  analytically the asymptotic behaviour
\be\la{1.15}
\Delta F(\lambda) \stackrel{\lambda\gg 1}{=} \frac{1}{2\pi}\l^{1/2}  \qquad\to\qquad 
\Delta q(\lambda) \stackrel{\lambda\gg 1}{=} -\frac{1}{16\pi}\lambda^{3/2}+\cdots\ , 
\ee
where we used (\ref{1.11}).
We will also confirm this result by a high-precision resummation and extrapolation analysis of the weak-coupling expansion 
by numerical methods including a conformal-mapping 
improved Pad\'e analysis.

 Let us note  that while the  coefficient in strong-coupling limit of $q$ in $\N=4 $ SYM case is positive, 
 $q^{\N=4}= { 1 \ov 96} \l^{3/2} +...$  (see \rf{1.3}),  it  was found  
 \ci{Beccaria:2021ksw} to be  negative in the $\N=2$ orbifold theory \rf{18}. The result in \rf{1.15} implies that it is also negative in the orientifold theory, 
  $q^{\orientif}= q^{\N=4}  +  \Delta q = ({ 1 \ov 96}  - {1\ov 16\pi} )  \l^{3/2} +...\approx -0.00948  \l^{3/2} +... $.
  It would be interesting  to  understand the reason for this 
  sign change on the dual string theory side where $q$  should be expressed in terms of the string partition function on the disc with one handle.

\

The rest of the   paper is organised as follows. 
In section 2  we shall describe  the localization matrix model representation for the expectation  value of the  $\frac{1}{2}$-BPS 
 Wilson loop in the orientifold   gauge theory. We  shall then   present the derivation of the relation \rf{1.11}  between the  coefficient $\Delta q$ 
 of the $1/N^2$ term 
 in the ratio of  the orientifold and $\N=4$ SYM Wilson loops  and the large $N$ limit $\Delta F(\l)$  of the  difference  of the corresponding 
  free  energies. This reduces the problem of determining the strong coupling limit of $\Delta q$ to  finding that of $\Delta F$. 
  
 In section 3  we shall  first study  $\Delta F(\l)$  at weak coupling   and then 
 find  its explicit representation \rf{3.22}, i.e. $ \Delta F =\ha \log \det (1 + M) =  -\sum_{n=1}^\infty {1\ov n} (-1)^n  \tr M^n$,   in terms of  an infinite-dimensional  matrix $M$  \rf{3.23}. 
 Each  $\tr M^n$ term in $\Delta F$  turns out to be  of  fixed order $n$ in products of $\zeta$-function values 
 when written in the weak-coupling expansion. In section 4 we  shall study the first two $\tr M$  and $\tr M^2$   terms   finding  that 
 at large $\l$   one has   $\tr M^n \sim \l^n$. 
 
 The  derivation of the  strong coupling   asymptotics  \rf{1.15} of the total $\Delta F$  implying $\Delta q \sim \l^{3/2}$ 
  is given in section 5. 
 In section 6   we shall  independently  test  this   $\Delta F \sim \l^{1/2} $  scaling  by two different numerical methods. 
 %We will first consider the approach based on Pade approximants using as an input many terms in the weak coupling expansion of $\Delta F$. 
 Some technical details are delegated to   appendices.

%\medskip
%The  methods used  here   %  have an interest in their own,  being, in principle,  
%may be, in principle, applicable to other similar $\mc N=2$  models. 
%One  candidate is the $\mc N=2$ superconformal $SU(N)$ gauge theory with $N_F=4$ fundamental   and  $N_A=2$ 
%   rank-2 antisymmetric  hypermultiplets.  In this case the dual string 
% theory is again a  IIB  orientifold of  AdS$_{5}\times S^{5}$
%where $S^{5}$ is modded out by a $\mathbb Z_{4}$ that mixes non-trivially the orbifold and orientifold 
%twists.%and originates from D3-D7 system \cite{Ennes:2000fu}.
%\footnote{In this model  the Weyl anomaly coefficients are 
%${\rm a}=\frac{N^{2}}{4}+\frac{N}{8}-\frac{5}{24}$ and ${\rm c}=\frac{N^{2}}{4}+\frac{N}{4}-\frac{1}{6}$.  The $\mc O(N)$ terms  in a  and c    should be possible  to  derive  on  the dual string theory  side as   discussed  in \cite{Blau:1999vz}.}

The  methods used  here  may    be 
applicable to  other similar $\mc N=2$  models. 
One  candidate is the $\mc N=2$ superconformal $SU(N)$ gauge theory with $N_F=4$ fundamental   and  $N_A=2$ 
   rank-2 antisymmetric  hypermultiplets.  In this case the dual string 
 theory   is  expected to be again a  IIB  orientifold of  AdS$_{5}\times S^{5}$
where $S^{5}$ is modded out by a $\mathbb Z_{4}$ that mixes non-trivially the orbifold and orientifold 
twists  \cite{Ennes:2000fu}.\footnote{In this model  the Weyl anomaly coefficients are 
${\rm a}=\frac{1}{4}N^2+\frac{1}{8}N-\frac{5}{24}$ and ${\rm c}=\frac{1}{4}N^2+\frac{1}{4}N-\frac{1}{6}$. 
 The $\mc O(N)$ terms  in a  and c    should be possible  to  derive  on  the dual string theory  side  as in 
 %v2
     \cite{Blau:1999vz} (see also  \cite{Aharony:1999rz,Naculich:2001xu}) using that here the background involves D7-branes wrapping AdS$_5$ and $S^3$ of  $S'^5$ with $R^2$terms in the effective 8-dimensional world-volume theory.}
However, the presence of fundamentals  means that  here the large $N$   expansion will go in powers of $1/N$  rather than 
$1/N^2$ and    thus will be different  in structure from \rf{1.5},\rf{1.6}.

\paragraph{Note added in v3:}
The exact value of the coefficient $C$ in (\ref{18}) in orbifold theory  was recently  found 
 in \ci{Beccaria:2022ypy}: \  $C = -{1\ov 32} $. 
%This value (-0.03125)
%is substantially larger than the estimate in (\ref{18}). This shows that the explored range of 
%lambda values is still too narrow to reach the asymptotic regime $\lambda \gg 1$ where
%the leading term dominates. This is consistent with the fact that in \red{[new paper]}
%logarithmic corrections are identified at strong coupling slowing
%down convergence.
Ref. \ci{Beccaria:2022ypy} 
 also found the exact values of  several leading coefficients in strong  coupling expansion of 
 free energy of the orientifold theory \rf{1.14}, in particular, 
  of the coefficient  $c_1$ 
of the leading $\l^{1/2}$  term: \ $c_1 = {1\ov 8} $.
In view  of \rf{1.11} this  determines also the exact value of the coefficient  of the leading $\l^{3/2}$ term in $\Delta q$.
Thus instead of  the  result \rf{1.15}  obtained below  by making a bold  assumption 
  that one can interchange the
large $\l$ expansion with computing   determinant of a infinite matrix
 one actually gets 
\be\la{5555}
\Delta F(\lambda) \stackrel{\lambda\gg 1}{=} \frac{1}{8} \l^{1/2} \qquad\to\qquad 
\Delta q(\lambda) \stackrel{\lambda\gg 1}{=} -\frac{1}{64}\lambda^{3/2}\ .
\ee
Note  that the coefficients   in \rf{1.15}  still give a good   approximation to  the ones  in 
\rf{5555};  this is consistent with the  fact  that \rf{1.15}   was also supported by  numerical  evaluation methods. 
%The exact value $1/8$ does not compare too badly with $1/(2\pi)$ in (\ref{1.15}) and we mention that 
The value $c_1 = {1\ov 8} $ is also   in agreement  with 
%improves further using the
 refined numerical estimates in \cite{Beccaria:2021hvt},
see discussion in section 4.2 of  \ci{Beccaria:2022ypy}.
There   value $c_1 = {1\ov 8} $  was also   established by a more precise numerical method in 
  recent paper \ci{Bobev:2022grf}.

\section{Matrix model representation  and $1/N^2$  correction to  Wilson loop}

The field content of the  orientifold theory 
is represented by  the adjoint $\mc N=2$ vector multiplet (gauge vector $A_{\mu}$, a complex scalar $\varphi$, and two Weyl fermions) and   rank-2  symmetric  and antisymmetric 
 hypermultiplets (each containing two complex scalars and two Weyl fermions). 
 The $\frac{1}{2}$-BPS Wilson loop is defined  in terms of the fields of the vector multiplet as 
\be
\la{2.1}
\mc W = \tr\mc P\,\exp\Big\{g_{_{\rm YM}}\oint\Big[i\,A_{\mu}(x) dx^{\mu}+\tfrac{1}{\sqrt 2}\big(\varphi(x)+\varphi^{+}(x)\big)\,ds\Big]\Big\},
\ee
where the contour $x^{\mu}(s)$  represents  a circle of unit radius. 

\def \sint {S_{\rm int}}

The supersymmetric  localization implies 
that  the  partition function  of this  gauge theory 
on a sphere $S^{4}$ of unit radius admits a 
representation in terms of an integral over the eigenvalues $\{m_{i}\}_{i=1}^{N}$ of a traceless hermitian  $ N \times  N$ matrix $m$ \cite{Pestun:2007rz}
\ba\la{2.2}
Z^{\orientif} &\equiv e^{-F^{\orientif} }= \int \mc Dm \,e^{-S(m)}\ , \qquad\qquad \\
 S(m) &= S_0(m)  + \sint(m)  \  , \ \ \ \ \quad  S_0= { 8 \pi^2 N\ov \l} \tr m^2   \ , \qquad \l = g^2_{_{\rm YM}}N \ , \la{221}\\
\mc Dm &\equiv  \prod_{i=1}^{N}dm_{i}\,\delta\big(\sum_{j}m_{j}\big)\,\big[\Delta(m)\big]^{2}\ , \qquad \qquad 
\Delta(m) = \prod_{i<j}(m_{i}-m_{j}) \ . \la{2.3}
\ea
In the case of $\N=4 $ SYM theory $\sint=0$  and the matrix model is Gaussian. 
As we shall discuss  the $1/N$ expansion, 
 we can neglect the instanton contribution  term in   $\sint$ so that\foot{  % (see,  e.g.,   \cite{Bourget:2018fhe}   and refs. there)
This is a specialization of the general analysis in \cite{Pestun:2007rz}. See also \cite{Bourget:2018fhe,Billo:2019fbi}  for  applications to other  $\N=2$ gauge theories.}
\ba
S_0(m) = \frac{8\pi^{2}N}{\lambda}\sum_{i}\, m_{i}^{2}\ , \qquad \qquad \sint(m) = 
\sum_{i,j}\Big[\log H(m_{i}+m_{j})-\log H(m_{i}-m_{j})\Big]\ ,  \la{2.4}
\ea
where   $H$ is expressed  in terms of the  Barnes G-function
\ba
H(x) &= \prod_{n=1}^{\infty}\left(1+\frac{x^{2}}{n^{2}}\right)^{n}\, e^{-\frac{x^{2}}{n}} = e^{-(1+\gamma_{\rm E})\,x^{2}}
\,{\rm G}(1+ix)\, {\rm G}(1-ix)\ , \la{24} \\
\log H(x)  & = \sum_{n=1}^{\infty}\frac{(-1)^{n}}{n+1}\zeta_{2n+1}\,x^{2n+2} \ . \la{2.6}
\ea
Here  and below  the constants   $\zeta_{2n+1}\equiv \zeta(2n +1)$   are the Riemann $\zeta$-function values. 

%Denoting $\mc Dm = \prod_{i=1}^{N}dm_{i}\,\delta(\sum_{i}m_{i})\,\Delta(m)^{2}$,
The normalized  expectation value of the Wilson loop (\ref{2.1}) can  be computed as the   matrix model average
of $\tr e^{2\pi m}$, i.e. 
\be
\vev{\mc W}^{\orientif} = \frac{\int\mc Dm\,e^{-S(m)}\,\tr e^{2\pi m}}
{\int\mc Dm\,e^{-S(m)}}= {1 \ov Z^{\orientif} }
{\int\mc Dm\,e^{-S(m)}\,\sum_{i}e^{2\pi m_{i}}}
\ .\la{2.7}
\ee
In the leading planar  approximation  $\sint$ is effectively 
suppressed  and thus we get  the same result as in the $\N=4$  SYM: 
$\vev{\mc W}^{\orientif} = \vev{\mc W}^{\N=4}  + \mc O({1\ov N}) $. 
Here we will be interested in the leading non-planar correction in the ratio  (cf. \rf{1.2},\rf{1.10})
\be \la{01}
\frac{\vev{\mc W}^{\orientif}}{\vev{\mc W}^{\N=4}}= 1 +\frac{1}{N^{2}}\,\Delta q(\lambda)+\mc O\Big(\frac{1}{N^{4}}\Big) \ . 
\ee
Let us show that $\Delta q(\lambda)$ can be represented as a $\l$ derivative \rf{1.11} of the difference of the free energies \rf{1.12}. 
The Wilson loop ratio \rf{01} may be  represented in general as 
\be \la{02} 
\frac{\vev{\mc W}^{\orientif}}{\vev{\mc W}^{\N=4}} =
\frac{\vev{e^{-S_{\rm int}} \, \tr e^{2\pi m}  }_0    }{ \vev{e^{-S_{\rm int}}}_0\, \vev{\tr e^{2\pi m}}_0\, } \ , \ \ \qquad \qquad 
  \vev{...}_0\equiv \int Dm\,e^{-S_0(m)}... \ ,  \ \ \    \vev{1}_0=1 \ , 
\ee
where $Dm$ is the normalized   measure, i.e. $\int Dm\,e^{-S_0(m)}... = {   \int \mc D m\,e^{-S_0(m)}...\ov  \int \mc D m\,e^{-S_0(m)}}   $. Then  $ \vev{\mc W}^{\N=4}= \vev{\tr e^{2\pi m} }_0$   and 
 \be \la{03}
  \vev{ e^{-S_{\rm int}}}_0={Z^{\orientif}\ov Z^{\N=4}}= e^{-\Delta F} \ , \ \ \ \ \ \ \ \ \ 
 \Delta F(\l; N) = F^{\orientif}-  F^{\N=4}
 \ . \ee 
% \vev{\mc W}^{\N=4}= \frac{\vev{\tr e^{2\pi m} }_0}{ \vev{1}_0}, \ \   $
At large $N$  the correlators in \rf{02}  factorize  and the ratio  goes  to 1. 
The  non-planar correction is given by   the large $N$ limit of the 
``connected'' part of $\vev{e^{-S_{\rm int}} \, \tr e^{2\pi m}  }_0 $.
The leading $N\to \infty$ contribution should  come from  the first non-trivial term in the 
expansion of  $\tr e^{2\pi m} $ in powers of $m$ \  ($\tr 1 = N, \ \tr m=0$):
\be\la{213}
 \vev{e^{-S_{\rm int}} \, \tr e^{2\pi m}  }_0 = N \vev{e^{-S_{\rm int}} }_0 + 2\pi^2 \vev{e^{-S_{\rm int}} \, \tr m^2 }_0 + ... \ .\ee
%where we used that. 
The insertion of a factor of $\tr m^2$   is the same as   the insertion of the free action in \rf{221}
and   thus it can  be obtained by differentiating the partition   function \rf{2.2} over $\l$.   
%MB-referee
As shown in Appendix~\ref{app-Dq}, taking the large $N$ limit we then find that $\Delta q$ in \rf{01} can be represented as in \rf{1.11}, i.e.
%\foot{One  can show  directly  that correction coming from the next term in the 
%%v2
% expansion in \rf{213} is subleading at large $N$. Its 
% contribution to $\Delta q$  may be written as 
%\be\no
%\lim_{N\to\infty}\frac{6\vev{\tr m^{2}}_0 \Big[\vev{\tr m^{2}}_0   \vev{e^{-S_{\rm int}}}_0-\vev{\tr m^{2}e^{-S_{\rm int}}}_0\Big] +N\Big[\vev{\tr m^{4}e^{-S_{\rm int}}}_0-\vev{\tr m^{4}}_0\vev{e^{-S_{\rm int}}}_0\Big]}
%{\vev{e^{-S_{\rm int}}}_0}.
%\ee
%Here the two square  brackets  in the  numerator   represent 
% two connected correlators  that go like $1/N^{2}$ and  thus give  vanishing  contributions for large $N$.
%}
\be\la{3.6}
 \Delta q = -\frac{\lambda^{2}}{4}\lim_{N\to\infty}\frac{\partial}{\partial\lambda}\Delta F(\lambda; N) 
 = -\frac{\lambda^{2}}{4}\frac{d}{d\lambda}\Delta F(\lambda) \ , \qquad \ \ \ \ 
 \Delta F(\lambda) = \lim_{N\to\infty}\Delta F(\lambda; N) \ .
\ee
%%%%%%%%%%%%%%%%%%%%%%%%%%%%%%
This is  essentially the same   relation as  was  observed to hold  in the $\N=2$ orbifold model in \cite{Beccaria:2021ksw}
(up to factor of 2  due to the $SU(N) \times SU(N)$  instead of $SU(N)$ gauge group).

%In what follows, we shall often omit the ``{\orientif}'' label.

\section{Large $N$ limit of free energy  difference $\Delta F= F^{\orientif}-  F^{\N=4} $}
%:  weak coupling expansion  and   explicit representation}  %and  non-planar correction to the  Wilson loop}

Since the leading non-planar correction to the Wilson loop 
can be expressed \rf{3.6}  in terms of $\Delta F(\l)$, 
   in what follows we shall concentrate on the study of  its structure both at weak and strong coupling. 
Redefining  the matrix model variable $m \to a$  as 
\be
a = \sqrt{\frac{8\pi^{2}N}{\lambda}}\,m \ ,
\ee
we can represent  $ \Delta F(\lambda; N) $ in \rf{03} as 
% of the  free energy  energies of the  orientifold theory  and the  $\mc N=4$ SYM as 
\be
\la{3.2}
e^{-\Delta F(\lambda; N)} =  \int Da\,e^{-S_{\rm int}(a)}\,e^{-\tr a^{2}}\ ,
\ee
where $Da$ is the standard   integration 
measure  for the  traceless matrix $a$, normalized so that  $\int Da\,  e^{-\tr a^{2}}=1$.
This   measure is same as  in \rf{2.3}  when written in terms of the eigenvalues 
and dropping  the ``angular'' part  that cancels in expectation values  of   relevant correlators (functions of traces of matrix $a$).

Using \rf{2.4},\rf{2.6}   we can write  $S_{\rm int}$ in  \rf{2.4}  as a weak coupling expansion
%The genuine $\mc N=2$ part  of  the corresponding action is 
 % $S_{\rm int}= \sum_{i,j}\Big[\log H(m_{i}+m_{j})-\log H(m_{i}-m_{j})\Big] $   that can be written,  using \rf{2.6}, as   %and it takes the form 
\ba
S_{\rm int}(a)  %&= \sum_{i,j}\Big[\log H(m_{i}+m_{j})-\log H(m_{i}-m_{j})\Big] \no \\
&= 2\,\sum_{n=1}^{\infty}\Big(\frac{\hat \lambda}{N}\Big)^{n+1}\,\frac{(-1)^{n}}{n+1}\zeta_{2n+1}\,\sum_{p=0}^{n}\binom{2n+2}{2p+1}\,\tr a^{2p+1}\,\tr a^{2n-2p+1} \  , \qquad \qquad  \hat\lambda \equiv \frac{\lambda}{8\pi^{2}}. \la{3.3}
\ea

\subsection{Weak coupling expansion}
\la{sec:dF-weak}

The weak coupling ($\l \ll 1$) expansion of $\Delta F$ in (\ref{3.2}) is easily worked out  by expanding $e^{-S_{\rm int}}$ and doing the Gaussian integrations. The result has a finite limit for $N\to \infty$ since the leading 
$N^{2}$ terms present in both the $\mc N=4$ SYM and $\mc N=2$ partition functions cancel out  in $\Delta F$
as a manifestation of the  planar equivalence of the two models. 
For the  leading large $N$ contribution $\Delta F(\lambda)$ defined in   \rf{3.6} % (appearing in   \rf{1.11}) 
%\be\la{3.4}
%\Delta F(\lambda) = \lim_{N\to\infty}\Delta F(\lambda; N) \ , 
%\ee
we obtain the following expansion (\cf  (\ref{1.13}))
\ba
&\Delta  F(\lambda) = 5 \zeta _5 \hat{\lambda }^3-\tfrac{105}{2} \zeta _7 \hat{\lambda 
}^4+441 \zeta _9 \hat{\lambda }^5-(25 \zeta _5^2+3465 \zeta _{11}) 
\hat{\lambda }^6+\big(525 \zeta _5 \zeta _7+\tfrac{3.6355}{8} \zeta _{13}\big) 
\hat{\lambda }^7  \nonumber \\
&-\big(\tfrac{22785 }{8}\zeta _7^2+\tfrac{8505 }{2}\zeta _5 \zeta _9+\tfrac{6441435}{32} \zeta _{15}\big) \hat{\lambda 
}^8+\big(\tfrac{500 }{3}\zeta _5^3+\tfrac{94815 }{2}\zeta _7 \zeta _9+\tfrac{63525 }{2}\zeta _5 \zeta _{11}  +\tfrac{12167155 }{8}\zeta _{17}\big) \hat{\lambda }^9
 \lp
- \big(5250 \zeta _5^2 \zeta _7+201852 \zeta 
_9^2+\tfrac{724185 }{2}\zeta _7 \zeta _{11}+\tfrac{920205 }{4}\zeta _5 \zeta _{13}
+\tfrac{91869921 }{8}\zeta _{19}\big) \hat{\lambda }^{10}+\cdots\ . 
%,\qquad \hat\lambda \equiv \frac{\lambda}{8\pi^{2}}. 
\la{3.5}
\ea
%%%%%%%%%%%%%%%%%%%%%%%%%%%%%%%%%%%%%%%
\iffa
As in the orbifold theory  case \cite{Beccaria:2021ksw}, it turns out that there  is a non-trivial relation \rf{1.11} 
 between the function $\Delta q(\lambda)$ in  \rf{1.2},(\ref{1.10})
and $\Delta F(\lambda)$ defined in  (\ref{1.12}),\rf{3.4}
 %\red{we don't have a proof, but only strong evidence, as in orbifold case. Maybe we can add these evidence that comes from the fact that the first  correction to the Wilson loop expectation is $\sim I_{1}$ etc.}}
\be\la{3.6}
\Delta q(\lambda) = -\frac{\lambda^{2}}{4}\frac{d}{d\lambda}\Delta F (\l) \ . 
\ee
\fi
%%%%%%%%%%%%%%%%%%%%%%%%%%%%%
%The strong evidence  for this relation  
A  check of the   general relation  \rf{3.6}  may be given by 
%comes  from 
the 
direct  comparison of the independent weak-coupling expansions   for $\Delta F$ and $\Delta q$  (see Appendix \ref{app-Dq}).
%  and  also 
%  and also  from the appearance of Bessel-function  $ I_{1}$ factors  (see
%  the  discussion in \cite{Beccaria:2021ksw}).
%Indeed, 
%It can be verified  by  working out  directly the 
%From (\ref{3.5}), one obtains the 
The weak coupling expansion of $\Delta q(\lambda)= q^{\orientif}(\lambda)-  q^{\N=4}(\lambda)$ is found to be 
\ba
\frac{1}{4\pi^{2}}\,\Delta q(\lambda) &=\te  -\frac{15}{2} \zeta_{5} \hat{\lambda }^4+105 \zeta_{7}\hat{\lambda 
}^5-\frac{2205}{2} \zeta_{9} \hat{\lambda}^6 +(75 \zeta_{5}^2+10395 \zeta_{11}) \hat{\lambda }^7   \lp\te \quad 
-\big(\frac{3675}{2} \zeta_{5} \zeta_{7}+\frac{1486485}{16} \zeta_{13}\big) \hat{\lambda }^8
+\big(\frac{22785 
\zeta_{7}^2}{2}+17010 \zeta_{5} \zeta_{9}+\frac{6441435 \zeta_{15}}{8}\big) \hat{\lambda }^9+\cdots\ , \la{3.7}
\ea
which  is  indeed consistent    with    \rf{3.5}  and \rf{3.6}.

\subsection{Explicit  representation for  $\Delta F$}

It is possible   to derive a remarkable   closed  expression  for $\Delta F(\lambda)$ in \rf{3.6} as a  $\log \det $  of an infinite-dimensional matrix. 

Let us start with representing   $ S_{\rm int}(a) $  in \rf{3.3}  as  an  infinite double 
sum of $1\ov \sqrt{N}$-normalized  traces of odd powers of  the matrix $a$ with coefficients $C_{ij}$ that depend only on $\l$ % we can write 
  \ba \la{37}
  S_{\rm int}(a)&= \sum^\infty_{i,j=1} C_{ij}(\l) \  \tr  \Big( {a\ov \sqrt{N} }\Big)^{2i+1}  \, \tr \Big( {a\ov \sqrt{N} }\Big)^{2j+1}  \ , \\
\la{3.15}
C_{ij}(\l)  &=  4\,\hat\lambda^{i+j+1}\,(-1)^{i+j}\,\zeta_{2i+2j+1}\frac{\Gamma(2i+2j+2 )}{\Gamma(2i+2)\, \Gamma(2j+2)} \ .
\ea
Next, let us  define the generating function
  % Let us  consider    the  following generating function ($\eta=(\eta_1, \eta_2, ...)$)
\be
X(\eta)  %\equiv  X(\eta_{1}, \eta_{2}, \eta_{3}, \dots)
 =\int Da\,  e^{-\tr a^{2}} \ e^{V(\eta, a)  }\ , \qquad \qquad 
 V(\eta, a) = \sum_{k=1}^\infty \eta_{k}\,  \tr \Big( {a\ov \sqrt{N} }\Big)^{2k+1} \ .\la{36}
 % \exp\Big[\frac{\eta_{1}}{N^{3/2}}\tr a^{3}+\frac{\eta_{2}}{N^{5/2}}\tr a^{5}+\frac{\eta_{3}}{N^{7/2}}\tr a^{7}+\cdots\Big]\,e^{-\tr a^{2}}.
\ee
Using \rf{37}  can then represent $e^{-S_{\rm int}(a)}$   and thus  the integral  over $a$  in \rf{3.2} as\foot{Here and below we assume summation over repeated   indices $i,j$.} 
\ba \la{39}
 e^{-S_{\rm int}(a)} &=e^{-  \mathscr{D}(\lambda)} \, e^{V(\eta, a)}\Big|_{\eta=0} \ , 
 \qquad \qquad      e^{-\Delta F(\lambda; N)} =  e^{-  \mathscr{D}(\lambda)} \,  X(\eta) \Big|_{\eta=0}  \ ,         \\
\la{3.14}
\mathscr{D}(\lambda) &\equiv  C_{ij}(\lambda)    \frac{\partial}{\partial\eta_{i}}\frac{\partial}{\partial\eta_{j}}  \ .
% \partial_i\partial_{j}, \qquad \partial_{i} = \frac{\partial}{\partial\eta_{i}}.
\ea
The large $N$ limit of $\Delta F(\lambda; N)$  is thus directly related to that of $X(\eta)$. 

Expanding $e^V$ in \rf{36}  in powers of $a$, computing the Gaussian integrals over $a$, and then rearranging the result  back into the exponential form   gives the following expression 
for the   leading large $N$ part of $X(\eta)$
\be\la{311}
X(\eta) = e^{Q(\eta)}\,\Big[1+\mc O\big( {1\ov N}\big)\Big]\ ,
\ee
where $Q(\eta)$ is the following    quadratic form
\ba\la{3.11}
Q(\eta) &\equiv  \te Q_{ij}\eta_{i}\eta_{j} = \frac{3}{16} \eta _1^2+\frac{15}{16} \eta _1 \eta _2+\frac{5}{4} \eta_2^2
+\frac{63}{32} \eta _1 \eta _3+\frac{175}{32} \eta _2 \eta _3+\frac{1575}{256} \eta _3^2+\cdots.
\ea
The  closed form of the infinite matrix  $Q_{ij}$ can be found from the results in \cite{Beccaria:2020hgy} 
\be\la{333}
Q_{ij} =\frac{1}{\pi} \frac{2^{i+j}\,i\,j\,\Gamma(i+\frac{3}{2})\,\Gamma(j+\frac{3}{2})}{(i+j+1)\,\Gamma(i+2)\,\Gamma(j+2)}\ . 
\ee
As a result, we get  from \rf{39},\rf{311}
\be
\la{3.13}
e^{-\Delta F(\l) } =    \lim_{N\to\infty}   e^{-\Delta F(\lambda; N)}  =  e^{-\mathscr{D}(\lambda)}\,e^{Q(\eta)} \Big|_{{\eta}=0}
\ . 
\ee
To evaluate (\ref{3.13}), let us  first change the  variables as  $\eta=Q^{-1/2}x$ so that $  Q_{ij} \eta_i\eta_j   = x_i x_i $  and 
%We have also 
\be
\mathscr{D} = C_{ij}\frac{\partial}{\partial\eta_{i}}\frac{\partial}{\partial\eta_{j}} = 
(Q^{1/2}CQ^{1/2})_{ij}\frac{\partial}{\partial x_{i}}\frac{\partial}{\partial x_{j}}.
\ee
Introducing the  notation 
\be
\widetilde C = 4 Q^{1/2}\,  C\, Q^{1/2}\ , \qquad \qquad \la{3.16}
\widetilde M = 4CQ =  Q^{-1/2}\, \widetilde C\,  Q^{1/2} \ , 
\ee
we  then have (here $\partial_{i}=\frac{\partial}{\partial x_{i}}$, \ $\mc N=\rm const$)
\ba
e^{-\Delta F} &= 
e^{-\frac{1}{4} \widetilde C_{ij}\partial_{i}\partial_{j}}\,e^{x^{2}} \Big|_{x=0} = 
e^{-\frac{1}{4} \widetilde C_{ij}\partial_{i}\partial_{j}}\,\mc N\int d{y}\,e^{{x}\cdot{y}-\frac{1}{4}{y}^{2}} \Big|_{{x}=0} =
 \mc N\int d{y}\,e^{-\frac{1}{4} {y}\widetilde C{y}-\frac{1}{4}{y}^{2}}\lp
= [\det(1+\widetilde C)]^{-1/2} = [\det(1+\widetilde{M})]^{-1/2} \ . \la{3.17}
\ea
Here we used  an auxiliary Gaussian integral over $y_i$ and that 
$ e^{- A_{ij} \partial_i \partial_j    }\,e^{x\cdot y}\big|_{x=0}  = e^{-A_{ij}y_{i}y_{j}}$.

Thus  we   find   the  following exact representation for $\Delta F (\l)$ in terms of the infinite   matrix  $\widetilde M$
\be
\la{3.18}
\Delta F = \frac{1}{2}\log \det(1+\widetilde M) = \frac{1}{2}\tr \log(1+\widetilde M) = \frac{1}{2}\sum_{n=1}^{\infty}\frac{(-1)^{n+1}}{n}\,\tr \widetilde M^{n}.
\ee
Note that the  last equality in (\ref{3.18}) and the explicit form of $C_{ij}$ in (\ref{3.15})  imply 
that each power of $\widetilde M$  defined in \rf{3.16}  brings in one extra  factor  of the $\zeta_k$ constants. 
Explicitly, we have 
\ba
\la{3.19}
\tfrac{1}{2}  \tr \widetilde M = & 5 \zeta _5 \hat{\lambda }^3-\tfrac{105}{2} \zeta _7 \hat{\lambda 
}^4+441 \zeta _9 \hat{\lambda }^5-3465 \zeta _{11} \hat{\lambda 
}^6+\tfrac{3.6355}{8} \zeta _{13} \hat{\lambda }^7 \lp \qquad 
-\tfrac{6441435}{32} 
\zeta _{15} \hat{\lambda }^8+\tfrac{12167155}{8} \zeta _{17} 
\hat{\lambda }^9-\tfrac{91869921}{8} \zeta _{19} \hat{\lambda }^{10}+\cdots,  \\
%%%
-\tfrac{1}{2\times 2} \tr \widetilde M^{2} =& -25 \zeta _5^2 \hat{\lambda }^6+525 \zeta _5 \zeta _7 \hat{\lambda 
}^7-\big(\tfrac{22785 }{8}\zeta _7^2 +\tfrac{8505 }{2}\zeta _5 \zeta _9\big) 
\hat{\lambda }^8  \lp 
+\big(\tfrac{94815 }{2}\zeta _7 \zeta _9+\tfrac{63525 }{2}\zeta _5 \zeta _{11}\big) \hat{\lambda }^9-(201852 \zeta _9^2
+\tfrac{724185 }{2}\zeta _7 \zeta _{11}+\tfrac{920205 }{4}\zeta _5 \zeta _{13}\big) \hat{\lambda }^{10}+\cdots, \notag  \\
%%%
\tfrac{1}{2\times 3 }  \tr\widetilde M^{3} = &\tfrac{500}{3} \zeta _5^3 \hat{\lambda }^9-5250 (\zeta _5^2 \zeta _7) 
\hat{\lambda }^{10}+\cdots \ .  \notag 
\ea
That way  the weak-coupling expansion in \rf{3.18}  reproduces   the  $\zeta_k$, $\zeta_k\zeta_n$, $\zeta_k\zeta_n\zeta_m, ...$  terms in the expansion  of $\Delta F$ in (\ref{3.5}).\footnote{Keeping only a finite number of $\zeta_{k}$ constants 
 in  the matrix  $\widetilde M_{ij}$ and using the first equality in 
 (\ref{3.18}) gives immediately the 
resummation of all monomials involving those $\zeta_{k}$. 
For example, the terms with only $\zeta_{5}$ and $\zeta_{7}$ come from 
the expansion of the exact expression
$\Delta F_{\zeta_{5}, \zeta_{7}} = \frac{1}{2}\log(1+10 \zeta_{5}\hat{\lambda }^3 -105 \zeta_{7}\hat{\lambda }^4 -\frac{735}{4}\zeta_{7}^{2} \hat{\lambda }^8 )$,
and so on.
}

The explicit form of  the matrix  $\widetilde M$  in  \rf{3.16}  appearing in  (\ref{3.18}) is 
\be
\la{3.20}
\widetilde M_{ij} %= 4(QC)_{ij} 
= \frac{16}{\pi} \sum_{k=1}^{\infty}\hat\lambda^{k+j+1}\,(-1)^{k+j}\,\zeta_{2k+2j+1}
\,  \frac{2^{i+k}\,i\,k\,\Gamma(i+\frac{3}{2})\,\Gamma(k+\frac{3}{2})}{(i+k+1)\,\Gamma(i+2)\,\Gamma(k+2)}
\,\frac{\Gamma(2k+2j+2)}{\Gamma(2k+2)\ \Gamma(2j+2)}.
\ee
Motivated by the analysis in \cite{Beccaria:2020hgy}, let us introduce the following matrix
\ba
M_{ij} &= 8\,\sqrt{2i+1}\sqrt{2j+1}\,\sum_{k=0}^{\infty}\Big(\frac{\hat\lambda}{2}\Big)^{i+j+k+1} (-1)^{k}\,c_{i,j,k}\,\zeta_{2i+2j+2k+1}\ , \la{3.21} \\
c_{i,j,k} &= \sum_{m=0}^{k}\frac{\Gamma(2i+2j+2k+2)}{\Gamma(m+1)\, \Gamma(2i+m+2)\, \Gamma(k-m+1)\, \Gamma(2j+k-m+2)}\ . \la{321}
\ea
Remarkably,  
%one can check directly (to high order in  small $\l$ expansion) 
$\widetilde M$ in \rf{3.20} and $M$ in \rf{3.21}   
%have 
%the same traces. %This  implies   that $\widetilde M$   and $M$   should  have the same  eigenvalues. 
%Indeed, 
 %and thus should be
     happen to be  related by a similarity transformation, 
\be    M = U^{-1} \widetilde M U \ , \la{222}  \ee 
where% (see Appendix \ref{app-sim})
\foot{Notice that $U$ is a  lower triangular  matrix due to the argument of the first  $\Gamma$ function in the denominator being non-positive integer for $j>i$.}
\be\la{322}
U_{ij} = \frac{(-1)^{1-j}2^{1-i}\sqrt{1+2j}\,\Gamma(2+2i)}{\sqrt 3\,\Gamma(1+i-j)\ \Gamma(2+i+j)} \ . \ee
%While  it  would be  important to find the  exact form of  the matrix $U$ 
%    In what follows we assume  that the relation  \rf{322}   is valid   for  any value of $\l$. 
%Although we do not have  have a proof, \footnote{\red{can we prove or sharpen ???}} 
%we empirically observe that powers of $M$ and $\widetilde{M}$ have the traces as series in  $\lambda$, as can be tested at high perturbative order.
One can then  replace  $\widetilde M$ in (\ref{3.18}) by $M$,  getting 
% Hence, in the following we shall write the relation (\ref{3.18}) in terms of $M$ 
\be
\la{3.22}
\Delta F(\l)  = \frac{1}{2}\tr \log(1+M) = \frac{1}{2}\sum_{n=1}^{\infty}\frac{(-1)^{n+1}}{n}\,\tr M^{n}.
\ee
The advantage of this form of $\Delta F$  is that the matrix $M$ in \rf{3.21}  admits the following 
Bessel function representation
\ba
\la{3.23}
M_{ij}
 &= 8\,(-1)^{i+j}\,\sqrt{2i+1}\sqrt{2j+1}\,\int_{0}^{\infty}\frac{dt}{t}\frac{e^{2\pi\,t}}{(e^{2\pi\,t}-1)^{2}}\,J_{2i+1}(t\,\sqrt{\lambda})\,J_{2j+1}(t\,\sqrt{\lambda}) \ , 
\ea
which   will prove to be useful in  the analysis of the strong-coupling limit.

\section{Contributions to $\Delta F$ of finite degree in $\zeta$-function values}

The weak coupling expansion of $\Delta F$ in  (\ref{3.5}) can be  represented as    %filtered by writing 
\be\la{4.1}
\Delta F = \sum_{n=1}^{\infty}\Delta F^{(n)}\ , \qquad   \qquad \Delta F^{(n)} = \sum_{k_1, ..., k_n}   c_{k_1 ... k_n} (\l) \ \zeta_{k_1} ...\zeta_{k_n}=  \frac{(-1)^{n+1}}{n}\tr M^{n}\ , 
\ee
where $\Delta F^{(n)}$ is the total contribution of  terms that are  %monomials that are 
products of a fixed number $n$ of the Riemann $\zeta$-function values. Equivalently,   $ \Delta F^{(n)} $  represents the 
%This  corresponds to separating the 
contribution of %from the trace of the powers
the  $\tr M^{n}$ term in \rf{3.22} (cf.   (\ref{3.19})). 
% For  generic  value of the coupling  $\l$, this is a possible regrouping of the full $\Delta F$.

Here we will   study   $\Delta F^{(n)}$,  
 computing, in particular, its leading strong coupling asymptotic expansion. 
 We shall   focus in detail on the $n=1$ term % that we study in great detail in order to assess how (\ref{4.2}) arises and  what are its subleading corrections. 
and then discuss 
%We will also provide some non-trivial information on
 the $n=2$ one. % before turning to all $n\geq 1$. 
 We will find that  
\be
\la{4.2}
\Delta F^{(n)}(\lambda)  \stackrel{\lambda\gg 1}{=}   C_{n}\lambda^{n}+\cdots.
\ee
In  the next section 5 
 we will compute all the coefficients  $C_{n}$  in \rf{4.2} and  then 
 evaluate the  sum  of all $\Delta F^{(n)}$ thus  determining  the strong coupling asymptotics of $\Delta F$.
 % at leading large $\lambda$. Here, 

\medskip
 Defining  %the identity
\ba
\la{4.3}
G(t,t') &\equiv   8\,\sum_{i=1}^{\infty}(2i+1)\,J_{2i+1}(t)\,J_{2i+1}(t') 
= -\frac{4tt'}{t^{2}-t'^{2}}\Big[t\,J_{1}(t)\,J_{2}(t')-t'\,J_{2}(t)\,J_{1}(t')\Big]\ ,
\ea
which  has also  the following  integral form  \cite{Tracy:1993xj}
\be
\la{4.7}
G(t,t') = 2\,t\,t'\,\int_{0}^{1}du\,J_{2}(t\sqrt u)\,J_{2}(t'\sqrt u)\ , 
\ee
we may  use 
  the Bessel function representation of the matrix $M$ (\ref{3.23})  to represent  the traces of $M^n $ as the  iterated integrals
\ba
\tr M &= \int_{0}^{\infty}\widehat{dt}\,G(t\sql, t\sql), \qquad\qquad \qquad  \widehat{dt} = \frac{dt}{t}\,\frac{e^{2\pi\,t}}{(e^{2\pi\,t}-1)^{2}},  \la{4.4} \\
\tr M^{2} &= \int_{0}^{\infty}\widehat{dt}\,\int_{0}^{\infty}\widehat{dt'} \, G(t\sql, t'\sql)\,G(t'\sql, t\sql), \la{4.5} \\
\tr M^{3} &= \int_{0}^{\infty}\widehat{dt}\,\int_{0}^{\infty}\widehat{dt'} \,\int_{0}^{\infty}\widehat{dt''}\,  G(t\sql, t'\sql)\,G(t'\sql, t''\sql)\,G(t''\sql, t'\sql)\ , \ \  \ \  {\it etc.}\la{4.6} 
\ea
%and so on. For later use, let us  mention  the 
%MB-referee
We remark that (\ref{4.3}) coincides with the Tracy-Widom kernel \cite{Tracy:1993xj} upon the change of variables $t = \sqrt x$. 
It remains to be clarified  whether this is a coincidence or there is some deeper relation to eigenvalue statistics. This Bessel kernel also appears in the BES equation \cite{Beisert:2006ez}
and seems prevalent in integrable equations/models.

\subsection{Term linear in $\zeta_n$} % the $\zeta$-function values}

$\Delta F^{(1)}$ or  $\tr M$   is just a single integral (\ref{4.4})  and may be treated exactly.  Using \rf{4.3}  we have 
\ba\la{4.8}
\Delta F^{(1)} &= \frac{1}{2}\tr M = 4\,\int_{0}^{\infty}\frac{dt}{t}\frac{e^{2\pi\,t}}{(e^{2\pi\,t}-1)^{2}}\,\sum_{i=1}^{\infty}(2i+1)\,\big[J_{2i+1}(t\,\sqrt{\lambda})\big]^{2}.
\ea
From  the identity
\be
4\sum_{i=1}^{\infty}(2i+1)\,\big[J_{2i+1}(x)\big]^{2} = x^{2}\,\big[J_{0}(x)\big]^{2}+(x^{2}-4)\,\big[J_{1}(x)\big]^{2},
\ee
we obtain 
\ba
\Delta F^{(1)} &= \int_{0}^{\infty}\frac{dt\,e^{2\pi t}}{t(e^{2\pi t}-1)^{2}}\Big(t^{2}\lambda\,\big[J_{0}(t\,\sql)\big]^{2}+(t^{2}\lambda-4)\,\big[J_{1}(t\sql)\big]^{2}\Big) \nonumber \\
&= \frac{\lambda}{2\pi}\int_{0}^{\infty}\frac{dt}{e^{2\pi t}-1}\,\Big(\big[J_{0}(t\sql)\big]^{2}-\frac{8\,J_{0}(t\sql)J_{1}(t\sql)}{t\sql}
+\frac{(12-t^{2}\lambda)\,\big[J_{1}(t\sql)\big]^{2}}{t^{2}\lambda}\Big), \la{4.11}
\ea
where we used integration by parts. % in the second line. 
 This expression is exact and may be expanded at weak or strong coupling.

\paragraph{Weak coupling expansion:} Using  
\be
\la{4.12}
\int_{0}^{\infty}dt\,\frac{e^{2\pi t}}{(e^{2\pi t}-1)^{2}}t^{2p+1} = \,{(2p+1)!\ov (2\pi)^{2p+2}} \,\zeta_{2p+1} \ , 
\ee
 and expanding in $\l$  we recover from (\ref{4.11})  the first line in (\ref{3.19}). 
One can  find  the following all-order  result\footnote{This remarkably simple form of the coefficients follows from the relation
\be\notag
\big[J_{0}(t)\big]^{2}-\frac{8\,J_{0}(t)J_{1}(t)}{t}
+\frac{(12-t^{2})\,\big[J_{1}(t)\big]^{2}}{t^{2}} = \frac{5}{192}\,t^{4}\,{}_{1}F_{2}(\tfrac{7}{2}; 4,5; -t^{2}).
\ee
}
\be
\la{4.13}
\Delta F^{(1)} = \frac{4}{\pi}\sum_{k=2}^{\infty}(-8)^{k}\,\frac{(k-1)\,k\,(k+2)\,\Gamma(k+\frac{1}{2})\Gamma(k+\frac{3}{2})}{\big[\Gamma(k+3)\big]^{2}}\zeta_{2k+1}\hat\lambda^{k+1}.
\ee
By the standard ratio test this shows that the radius of convergence is $\pi^{2}$, as could  be expected. 
Indeed, $\l=\pi^2$  is the radius of convergence of perturbative expansion in $\mc N=4$ 
SYM theory in the planar limit (as suggested by the single-magnon dispersion relation, fixed by the superconformal symmetry \cite{Beisert:2004hm,Beisert:2006ez}, or  by the  
quantum algebraic curve approach  \cite{Gromov:2017blm}). The same  is expected to  apply also to the 
$\mc N = 2$ superconformal  theories % like  the orientifold one, 
(as was  first observed in the mass-deformed $\mc N=2^{*}$ theory \cite{Russo:2013qaa}, and recently  found also  %discovered 
in the orbifold theory case \cite{Beccaria:2021ksw}).

Expanding the exponentials in the integral in (\ref{4.11}) gives  an  alternative representation
in terms of  a sum of elliptic integrals
\be
\la{4.14}
\Delta F^{(1)} = 2\, \sum_{n=1}^{\infty}n\,\Big[-1-\frac{8\pi^{2}n^{2}-\lambda}{2\pi \lambda}\,\mathbb{E}\Big(-\frac{\lambda}{\pi^{2}n^{2}}\Big)
+\frac{8\pi^{2}n^{2}+7\lambda}{2\pi \lambda}\,\mathbb{K}\Big(-\frac{\lambda}{\pi^{2}n^{2}}\Big)\Big].
\ee
Here  one sees explicit singularities at $\lambda = -\pi^{2}n^{2}$ where the argument of $\mathbb K$  becomes  unity.

%%%%%%%%%%%%%%%%%%%%%%%%%%%%%%
\paragraph{Strong coupling expansion:}

The strong coupling (asymptotic) expansion of $\Delta F^{(1)}$ may be computed by Mellin transform methods \cite{zagier,flajolet}.
Defining  the Mellin transform  $ \widetilde{f}(s) = \int_0^\infty dx \,  x^{s-1}\,f(x)\,$  and considering  the convolution 
\be
\la{4.15}
(f\star g) (x)= \int_0^\infty dt \,  f(t\,x)\,g(t),
\ee
we have   $(\widetilde{f\star g})(s)  = \widetilde{f}(s)\,\widetilde{g}(1-s)$.
Let $\alpha< s< \beta$ be the fundamental strip of analyticity of $\widetilde{f}(s) $.
The asymptotic expansion of $f(x)$ for $x\to\infty$ is obtained by 
looking at the poles of $\widetilde{f}(s)$ in the region $s\ge \beta$. Then    the pole 
$
\frac{1}{(s-s_0)^N} $ in the Mellin transform  leads to  the term  $ \frac{(-1)^N}{(N-1)!}\,\frac{1}{x^{s_0}}\,\log^{N-1} x
$ in the original function. 
In our case, we can compare the right hand side of (\ref{4.11}) with (\ref{4.15}) as 
%\be
%\lambda^{-1}\,\tr QC = \frac{1}{4\pi}\int_{0}^{\infty}\frac{dt}{e^{2\pi t}-1}\,\Big[J_{0}(t\sql)^{2}-\frac{8\,J_{0}(t\sqrt\lambda)J_{1}(t\sqrt\lambda)}{t\sqrt\lambda}
%+\frac{(12-t^{2}\lambda)\,J_{1}(t\sqrt\lambda)^{2}}{t^{2}\lambda}\Big],
%\ee
\ba
x &= \sql, \qquad g(t) = \frac{1}{4\pi}\frac{1}{e^{2\pi t}-1}, \qquad 
f(t) = \big[J_{0}(t)\big]^{2}-\frac{8\,J_{0}(t)J_{1}(t)}{t}
+\frac{(12-t^{2})\,\big[J_{1}(t)\big]^{2}}{t^{2}}.
\ea
The Mellin transform is then 
\be
(\widetilde{f\star g})(s) = \frac{2^{-6+s} s (2+s) \csc ^2(\frac{\pi  s}{2}) \Gamma (2-s)\ \zeta 
(s)}{\big[\Gamma (1-\frac{s}{2})\big]^2\ \Gamma (2-\frac{s}{2})\ \Gamma 
(3-\frac{s}{2})},
\ee
and the asymptotic expansion at strong coupling can be extracted from the poles at $s=0, 1, 2, \dots$. This gives
\be
\la{4.18}
\Delta F^{(1)} = \frac{\lambda}{16\pi^{2}}-\frac{\sql}{2\pi^{2}}+\frac{1}{6}+\frac{\sql}{2\pi^{7/2}}
\sum_{p=1}^{\infty}
\frac{\Gamma(\frac{5}{2}+p)\Gamma(p-\frac{1}{2})\Gamma(p-\frac{3}{2})}{\Gamma(p)}
\frac{\zeta_{2p+1}}{\lambda^{p}}.
\ee
 The infinite sum in \rf{4.18}  has zero radius of convergence, with factorially divergent coefficients.\footnote{Let us note 
  that replacing the $\zeta$-values  by  the integral  using (\ref{4.12}) and  doing the sum, we obtain another representation
\be\notag
\Delta F^{(1)} = \frac{\lambda}{16\pi^{2}}-\frac{\sql}{2\pi^{2}}+\frac{1}{6}+\frac{2\lambda}{\pi^{3}}\int_{0}^{\infty}\frac{dt}{e^{2t\sql}-1}\Big[\mathbb{K}(t^{2})-(1+8t^{2})\,\mathbb{E}(t^{2})\Big].
\ee
This integral has a logarithmic singularity at $t=1$ on the $t$ integration contour, and so should be understood as an average above and below the cut.
}
The leading order  $\l$ term  corresponds to the $n=1$  case of %, linear in $\lambda$ agrees with 
  the general pattern (\ref{4.2}).

%\paragraph{Leading term at strong coupling:} 
The leading term in (\ref{4.18}) can be derived more directly. % in the following way.
We can  expand  the 
integrand in (\ref{4.4}) at large $\lambda$ and read off the coefficient of a suitable power of $\lambda$ from a convergent integral\foot{This procedure works for the leading order;  at subleading orders one gets 
divergent integrals 
%MB-referee
requiring a more careful treatment.} %In more details, we have 
\ba
\la{4.19}
\tr M &= \int_{0}^{\infty}\frac{dt}{t}\,\frac{e^{2\pi\,t}}{(e^{2\pi\,t}-1)^{2}}\,G(t\sql, t\sql) = 
\int_{0}^{\infty}\frac{dt}{t}\,\frac{e^{2\pi\,t/\sql}}{(e^{2\pi\,t/\sql}-1)^{2}}\,G(t, t) \lp 
= \frac{\lambda}{2\pi^{2}}\int_{0}^{\infty}\frac{dt}{t}\big[J_{2}(t)^{2}-J_{1}(t)J_{3}(t)\big] +\dots = \frac{\lambda}{8\pi^{2}}+\dots\ . 
\ea
As  $\Delta F^{(1)} = \frac{1}{2}\tr M$ (cf. \rf{4.8}), this  result is thus in agreement with (\ref{4.18}).

%%%%%%%%%%%%%%%%%%%%
\subsection{Term quadratic in  $\zeta_n$} % \ \ $\tr M^2$} %the $\zeta$-function values}

In the case of $\Delta F^{(2)} = -\frac{1}{4}\tr M^{2}$  in  \rf{4.1} 
 we can obtain an all-order weak coupling expansion in almost-closed form. 
 Although it is not as explicit as
(\ref{4.13}) for $\Delta F^{(1)}$, it may be used to generate a very large number of terms.
 Here we will  present the final result, with  details  given  in Appendix ~\ref{app:M2}.
Let us define the polynomials
\be
\la{4.20}
d_{\ell}(x) = (-1)^{\ell}\sum_{p=0}^{\ell}\frac{P_{p}^{(2,-2p-5)}(1-2x^{2})\ P_{\ell-p}^{(2,-2\ell + 2 p-5)}(1-2x^{2})}{4^{p+2}4^{\ell-p+2}\Gamma(p+3)\Gamma(p+4)\Gamma(\ell-p+3)\Gamma(\ell-p+4)},
\ee
where $P^{(\alpha, \beta)}_{n}(x)$ are Jacobi polynomials. We may write $d_\ell$  in the form 
\be
\la{4.21}
d_{\ell}(x)  = x^{\ell}\mathop{\sum_{m=m_{0}}}_{\Delta m = 2}^{\ell}a_{m}^{(\ell)}(x^{m}+x^{-m}),
\ee
where $m_{0} = 0/1$ if $\ell$ is even/odd and $m$ varies in steps of 2. 
The weak coupling expansion of $\tr M^{2}$  can then be  written in terms of sums  with 
 coefficients $a_{m}^{(\ell)}$ that are easily computed from \rf{4.20},(\ref{4.21})
\be
\la{4.22}
\tr M^{2} = 8\sum_{\ell=0}^{\infty}(2\pi)^{-12-2\ell}\,\lambda^{\ell+6}\,\sum^\ell_{m}a_{m}^{(\ell)} \Gamma(\ell+6+m)\Gamma(\ell+6-m)\, \zeta_{\ell+5+m}\, \zeta_{\ell+5-m}.
\ee

\paragraph{Leading term at strong coupling:} 
The expansion (\ref{4.22}) may not  be used directly at strong coupling. Nevertheless, we succeed in applying the manipulation 
we exploited in (\ref{4.19}). Indeed, we have
\be
\tr M^{2} = \lambda^{2}\int_{0}^{\infty}\int_{0}^{\infty}dt dt'\ \frac{[t' J_{1}(t')J_{2}(t)-t J_{1}(t)J_{2}(t')]^{2}}{\pi^{4}tt'(t^{2}-t'^{2})^{2}}+\dots
\ee
The integrand is symmetric so we write
\be
\int_{0}^{\infty}\int_{0}^{\infty}dt dt' f(t,t') = 2\int_{0}^{\infty}dt\int_{0}^{t}dt' f(t,t') = 
2\int_{0}^{\infty}dt\,t\,\int_{0}^{1}dx f(t,tx).
\ee
Doing first the integral over $t$, we get % the result
\ba
\la{4.25}
\tr M^{2} &= \lambda^{2}\int_{0}^{1}dx\, \tfrac{x(15-7x^{2}-7x^{4}+15 x^{6})-3(1-x^{2})^{2}(5+6x^{2}+5x^{4})\, \text{arctanh}\, x }{144\pi^{6}x^{5}} +\dots 
= \frac{\lambda^{2}}{192\pi^{4}}+\dots \ . 
\ea
This strong-coupling asymptotics 
follows again the   general  pattern (\ref{4.2}). 
A numerical test of this prediction will be discussed in section \ref{sec:pade}.

%%%%%%%%%%%%%%%%%%%%%%%%%%
\section{Strong coupling limit of $\Delta F$: analytic derivation}
\la{sec:sc}

Let us now generalize the  derivation  of strong-coupling limit  to the full $\Delta F$.
%of 
% $\Delta F^{(1)}$ and $\Delta F^{(2)}$ in (\ref{4.19})  and (\ref{4.25})  to  arbitrary  $\Delta F^{(n)}$. 
 The starting point will be  the explicit form of 
 %Let us first consider explicitly
   the  large $\l$  expansion of the matrix $M$ in \rf{3.23}.  It can be found 
%This can be computed 
 as in \rf{4.15}--\rf{4.18}   using  the  Mellin transform.
We have   % (\cf  (\ref{4.15}))
\ba
M_{ij} = &8(-1)^{i+j}\sqrt{(2i+1)(2j+1)}\, N_{ij} \ ,  \la{68}\\
N_{ij} \equiv & \sql\,(f\star g_{ij})(\sql)\ , \qquad \qquad 
f(t) = \frac{e^{2\pi t}}{(e^{2\pi t}-1)^{2}}\ ,\qquad \ \ g_{ij}(t) = \frac{1}{t}J_{2i+1}(t)J_{2j+1}(t).
\ea
Evaluating the Mellin transforms and taking residues, we get the asymptotic   expansion of $N_{ij}$ 
\ba
N_{ij} \stackrel{\lambda\gg 1}{=}    & \Big[\frac{\delta_{ij}}{i(i+1)(2i+1)}+\frac{\delta_{i+1,j}}{(i+1)(2i+1)(2i+3)}+\frac{\delta_{i,j+1}}{i(2i-1)(2i+1)}\Big]\,\frac{\lambda}{64\pi^{2}}\notag\\
& -\frac{\delta_{ij}}{24\,(2i+1)}+\frac{\zeta_3 }{2\pi^{2}}\,\cos\big(\pi(i-j)\big)\,  \frac{1}{\sql}+\cdots. \la{610}
\ea
Then the leading  strong-coupling part  of $M$  may be written as 
\ba
& \qquad \qquad  \qquad  M   \stackrel{\lambda\gg 1}{=}  {\l\ov 2 \pi^2}  \S  + ...\ , \la{58}  \\
& \S_{ij} = {1 \ov 4} (-1)^{i+j}\sqrt{\frac{2j+1}{2i+1}}\Big[\frac{\delta_{ij}}{i(i+1)}+\frac{\delta_{i+1,j}}{(i+1)(2i+3)}+\frac{\delta_{i,j+1}}{i(2i-1)}\Big]\ , \la{59}
\ea
where $\S$ is a symmetric three-diagonal infinite-dimensional matrix. 
As a result, we get 
\be
\la{5.1}
\tr M^{n}  \stackrel{\lambda\gg 1}{=}  b_{n}\,\Big(\frac{\lambda}{2\pi^{2}}\Big)^{n}+\cdots, \ \ \ \ \ \ \ \ \ 
b_{n} = \tr \S^{n} \ .
\ee
The explicit values of the coefficients 
 $b_n$  (related  to $C_n$ in  \rf{4.2} as $C_n = {(-1)^{n+1}\ov n (2 \pi^2)^n}    b_n $)
  are  given  in Appendix \ref{NEW}.

Remarkably, $\S$ in \rf{59}  is  essentially the same (up to 1/2) as the matrix appearing in Eq.~(2.7) of \cite{ikebe}. 
It follows from the analysis in  \cite{ikebe}  that in the infinite matrix limit the eigenvalues $\{\ss_{1}, \ss_{2}, \dots\}$ 
of $\S$ are 
\be
\ss_{k} = \frac{2}{\jj_{1,k}^{2}} \ ,\qquad \qquad   k=1, 2, ..., \la{510}
\ee 
where $\jj_{1,k}$ are the zeroes of the Bessel function $J_{1}(x)$.
\iffalse 
\foot{
%A25
This implies, in particular,  that the   
generating function   for the coefficients  $b_n=\tr \S^n $, i.e.  $
b(x)=\sum_{n=1}^{\infty} b_{n}x^{n-1} $ is given by 
$b(x) = \frac{1}{\sqrt{2x}}\frac{J_{2}(\sqrt{2x})}{J_{1}(\sqrt{2x})} $.}
\fi
Hence, we  get the following remarkable relation\footnote{This follows  from the Weierstrass infinite product representation of the Bessel function in terms of its zeroes:

$J_{\nu}(z) = \frac{(z/2)^{\nu}}{\Gamma(\nu+1)}\prod_{n=1}^{\infty}(1-\frac{z^{2}}{{\rm j}^2_{\nu,n}})$, see for instance section 15.41 in \cite{watson}.}
\be   \la{555}
 \det \big(1 + \frac{\lambda}{2\pi^{2} } \S  \big) = \prod_{k=1}^\infty \big( 1 +   \frac{\lambda}{\pi^{2} } { 1 \ov \jj^2_{1,k}} \big) 
=  \frac{2\pi}{i\sql}\,J_{1}\Big(\frac{i\sql}{\pi}\Big) = \frac{2\pi}{\sql}\,I_{1}\Big(\frac{\sql}{\pi}\Big) \ .
\ee 
As a result, we get for $\Delta F$ in   \rf{3.22} 
\ba
\la{5.10}
\Delta F  &= \frac{1}{2}\log\det (1+M)  \stackrel{\lambda\gg 1}{=}   \frac{1}{2} \log\det (1 + \frac{\lambda}{2\pi^{2} } \S  ) 
= \frac{1}{2}\log\bigg[\frac{2\pi}{\sql}\,I_{1}\Big(\frac{\sql}{\pi}\Big)\bigg]\ \stackrel{\lambda\gg 1}{=} \   \frac{\sql}{2\pi}+\cdots \ . 
\ea
%Combining \rf{5151} and \rf{5.9} 
%the final result  for the strong-coupling limit of $\Delta F$   is thus 
%\be
%\la{5.10}
%\Delta F(\lambda)  \stackrel{\lambda\gg 1}{=}  
% \frac{1}{2}\log\Big[\frac{2\pi}{\sql}\,I_{1}\Big(\frac{\sql}{\pi}\Big)\Big]\ \stackrel{\lambda\gg 1}{=} \   \frac{\sql}{2\pi}+\cdots \ . 
%\ee
Eq.\rf{5.10}  implies that $c_1$ in \rf{1.14}  is equal to $\frac{1}{2\pi}$.
Then  using \rf{1.11} we obtain  the following expression \rf{1.15}  for the strong-coupling limit of $\Delta q$  
\be
\Delta  q(\lambda) \stackrel{\lambda\gg 1}{=}  -\frac{\lambda^{3/2}}{16\pi}+\cdots. \la{5.11} 
\ee

\section{Numerical evaluation of $\Delta F$:  interpolation from small  to   large $\l$} % strong coupling}

In this final section we present various approaches to test the analytical result  (\ref{5.10}) for the strong coupling limit of $\Delta F$  by numerical methods.
We will first  consider  the approach   based on  Pad\'e approximants using as an  input many terms 
in  the weak coupling expansion
of $\Delta F$. 
Then, we will discuss a  method based on a direct evaluation of $\Delta F=\frac{1}{2}\tr\log(1+M)$ where the large $\lambda$ limit  of the  infinite matrix  $M$ is first  replaced
by its  finite-size  truncation. % of its leading order large $\lambda$ expansion.

\subsection{Pad\'e-conformal method}
\la{sec:pade}

We begin with the small $\lambda$ expansion of $\Delta F$:
\be
\Delta F(\tilde\lambda)=\sum_k c_k \tilde{\lambda}^k\ , \qquad \qquad \qquad \tilde{\lambda} \equiv  8\hat\lambda = \frac{\lambda}{\pi^2}\ . 
\label{6.1}
\ee
%where $\tilde{\lambda} = 8\hat\lambda = \frac{\lambda}{\pi^2}$. 
The  particular definition of  $\tilde \lambda$ is chosen so that the radius of convergence of the series in \rf{6.1}  is as close as possible to 1. 
This is helpful for the numerical analysis, as it avoids the appearance of very large or very small coefficients at high order. 

The technical goal is to extrapolate from  small   to 
large $\lambda$, starting from a {\it finite} number of terms in the weak coupling expansion.  
 Optimal and near-optimal methods for such an extrapolation have been 
analyzed recently in \cite{Costin:2019xql,Costin:2020hwg,Costin:2020pcj}. The key information is some knowledge, either analytic or numerical, of the singularity 
structure of the function $\Delta F(\tilde\lambda)$.  This information can be extracted numerically by suitable combinations of ratio tests, Pad\'e approximants, and conformal maps. 

The magnitude of the leading singularity is equal to the radius of convergence $R$, which can be found by a simple ratio test:
\be
\left| \frac{c_{k+1}}{c_k}\right| \to \frac{1}{R},  \qquad k\to \infty.
\label{6.2}
\ee
The convergence of this ratio of successive coefficients to the inverse radius can be accelerated using Richardson acceleration \cite{Bender-Orszag} (for example,  for the $\tr M$ case see the left hand panel of  Figure \ref{fig:trm1-ratio} below).

 This permits an extremely precise numerical estimate of the radius of convergence, 
if it is not known analytically. For $\Delta F=\frac{1}{2}{\rm tr}\log(1+M)$, we will see that the leading singularity is at  $\td \l \approx -1$, i.e. 
$\lambda\approx -\pi^2$. %, which motivates the definition of the rescaling $\tilde{\lambda}\equiv \frac{\lambda}{\pi^2}$.
By studying the subleading corrections to this ratio test limit one can determine the nature of the leading singularity, using Darboux's theorem, see Appendix \ref{app:darboux}.  
For this orientifold model the small $\lambda$ expansion indicates that the leading singularity is logarithmic (see the right hand panel of  Figure \ref{fig:trm1-ratio} below). This is consistent with the exact analytical structure of  individual $\tr M^{n}$ terms for finite $n$, see section \ref{sec:sc}.

A closely related method, which also yields information about the singularity structure is  based on the  use of a Pad\'e approximant \cite{Bender-Orszag,Baker}. Here one matches the finite number $K$ of terms of the expansion to the expansion of a ratio of polynomials $R_L$ and $Q_M$:
\be
\mathcal P_{[L,M]}\Big[\sum_{k}^{K} c_k  \tilde{\lambda}^k\Big]=\frac{R_L(\tilde\lambda)}{Q_M(\tilde\lambda)}+O(\tilde\lambda^{K+1})\ . 
\label{6.3}
\ee
Since it is an approximation in terms of rational functions, Pad\'e only has poles as singularities, which are the zeros of the denominator polynomial $Q_M$. 
If the truncated series is that of a function with branch point singularities, then Pad\'e produces arcs of poles accumulating at the branch points.\footnote{There is a deep connection to 
electrostatics and potential theory, whereby (in this interpretation it is easiest to consider an expansion about infinity instead of about zero) in the $K\to\infty$ limit a Pad\'e approximation
produces lines of poles that form a capacitor having minimal capacitance \cite{Stahl,Costin:2020pcj}.} The practical implication of this is that if one has enough 
expansion terms one can frequently distinguish between an isolated pole and a branch point simply by looking at the poles of a Pad\'e approximant. 
Indeed, the left panel of 
Figure \ref{fig:trm1-pade-poles}  shows a line of Pad\'e poles accumulating to the branch point at $\tilde\lambda=-1$. 

However, this reveals a fundamental problem with Pad\'e, because these accumulating poles, which are trying to represent a branch cut, obscure possible higher 
singularities which may be physical. This problem can be resolved by making a conformal map before making the Pad\'e approximation \cite{Costin:2019xql,Costin:2020hwg,Costin:2020pcj}. 
Based on the leading branch cut $(\infty, -1]$ on the negative real $\tilde{\lambda}$ axis, as suggested by the Pad\'e approximation in this case (see the left hand panel of Figure \ref{fig:trm1-pade-poles}), one maps the expansion into the unit disk $|z|\leq 1$:
\be
\la{6.4}
z=\frac{\sqrt{1+\tilde\lambda}-1}{\sqrt{1+\tilde\lambda}+1}, \qquad \qquad \tilde\lambda=\frac{4z}{(1-z)^2}.
\ee
We re-expand $\Delta F\left(\frac{4z}{(1-z)^2}\right)$ in powers of $z$ to the same order $K$, and {\bf then} construct a Pad\'e approximant in terms of $z$.\footnote{As a technical comment: 
when dealing with high order Pad\'e approximants, numerical instabilities can arise due to close zeros and poles, also associated with large coefficients of the Pad\'e polynomials. 
This instability can be ameliorated by converting the Pad\'e representation to a partial fraction expansion, which  in principle is equivalent but in practice is more stable numerically.}  
Inside the unit disk this expansion is convergent by construction, but further singularities along the line $\tilde\lambda\in (\infty, -1]$ will appear as singularities on the unit circle. 
If these are branch points they will appear as the accumulation points of arcs of Pad\'e poles. 

The advantage of the conformal map is that collinear singularities in the $\tilde\lambda$ 
plane (which may be hidden under a line of accumulating poles) are separated to different points on the unit circle. See for example the right panel of 
Figure  \ref{fig:trm1-pade-poles}, which shows the 
leading singularity at $z=-1$, the conformal map image of $\tilde\lambda=-1$, but also clearly shows further singularities at the conformal map images of $\tilde\lambda=-4$, at $\tilde\lambda=-9$, and so on. 
This numerical evidence suggests that the singularities are:
\be
\la{6.5}
{\rm singularites}\big(\Delta F(\lambda)\big)=-l^2\, \pi^2\ , \qquad\qquad  l=1, 2, 3, \dots\ .
\ee
The source of these singularities can be understood analytically from the study of $\tr M^{n}$ for finite $n$, and the singularity structure appears to be inherited by ${\rm tr}\,\log(1+M)$.

A further advantage of the conformal map is that it enhances the precision of the subsequent Pad\'e extrapolation. To construct the Pad\'e-conformal extrapolation\footnote{This was applied to the 
Borel transform function in \cite{Costin:2019xql,Costin:2020hwg}, but it can also be applied to any series with a finite radius of convergence \cite{Costin:2020pcj}.} we make a Pad\'e approximant 
in terms of $z$ and then evaluate it on the inverse map in (\ref{6.4}). This introduces square roots; thus we are representing the function not just by rational approximations, but in a much 
wider class of functions. For branch point singularities the increase in precision can be quantified precisely using the asymptotics of orthogonal polynomials \cite{Costin:2020hwg} and is 
quite dramatic, as is illustrated %here in this example
 in Figures \ref{fig:trm1-extrap} and \ref{fig:trlog1pm-extrap} below.

\subsubsection{Example:  $\tr M$}

To illustrate this Pad\'e-conformal extrapolation technique, we first consider the expansions of $\tr M$ and $\tr M^{2}$, for which we can compare with analytic results found in section 4.  
But we stress that the power of this method is in cases when such analytic comparisons are not available, and one is only presented with a truncated series, and possibly some physical 
intuition about the singularity structure. For $\tr M$ we have the exact expansion (\cf (\ref{4.13}))
\be
\la{6.6}
\tr M =\sum_{k=2}^\infty \frac{(-1)^k (k-1) k (k+2) \,  \zeta_{2 k+1}\, 
 \Gamma \left(k+\frac{1}{2}\right) \Gamma \left(k+\frac{3}{2}\right)}{\pi \big[ \Gamma (k+3)\big]^2} \, \tilde\lambda^{k+1}\ .
\ee
%%%%%%%%%%%%%%%%%%%%%%%%%%%%%%%%
\begin{figure}[htb]
\begin{center}
\includegraphics[scale=.6, width=0.45\textwidth]{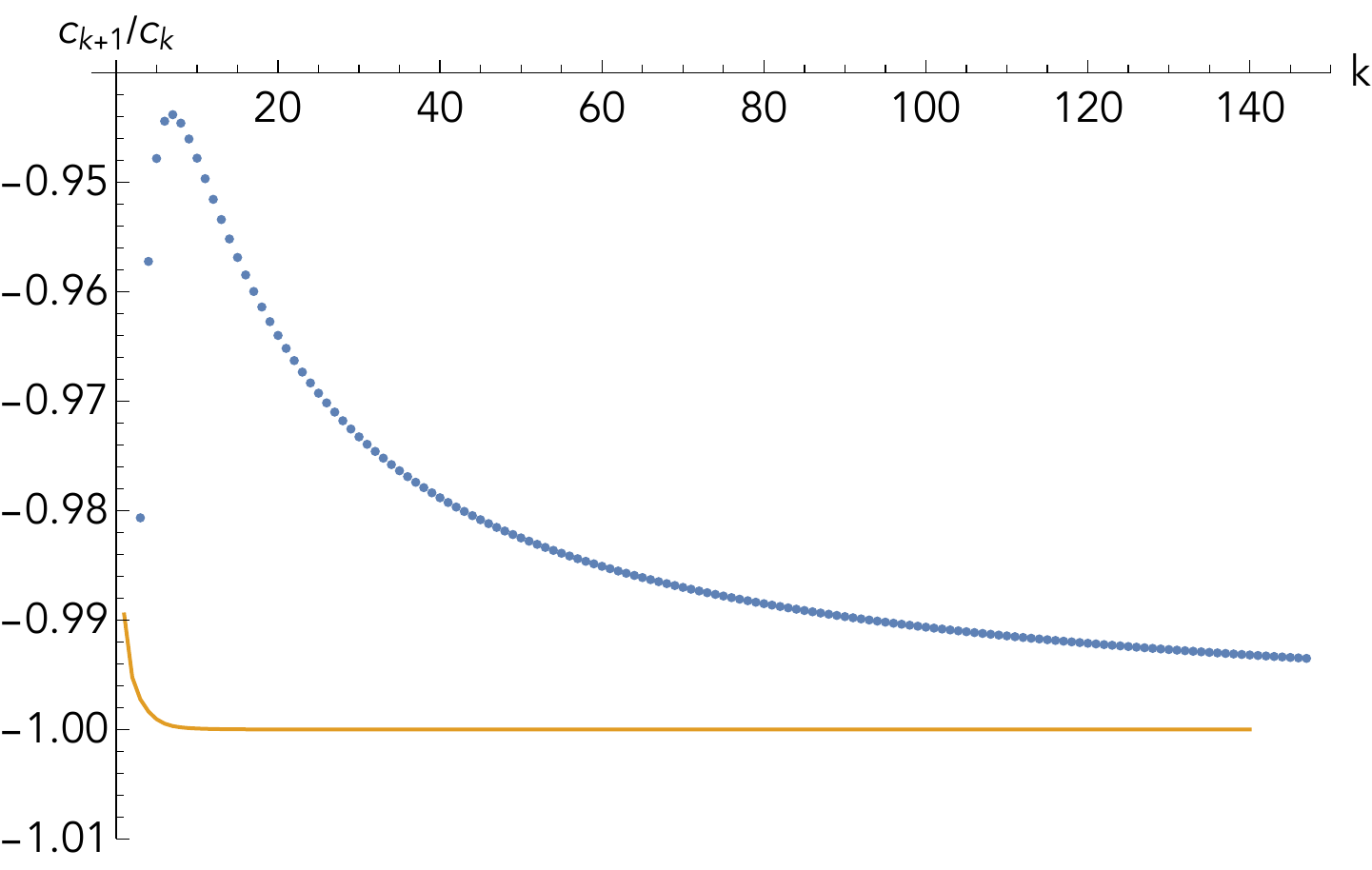}
\includegraphics[scale=.6, width=0.45\textwidth]{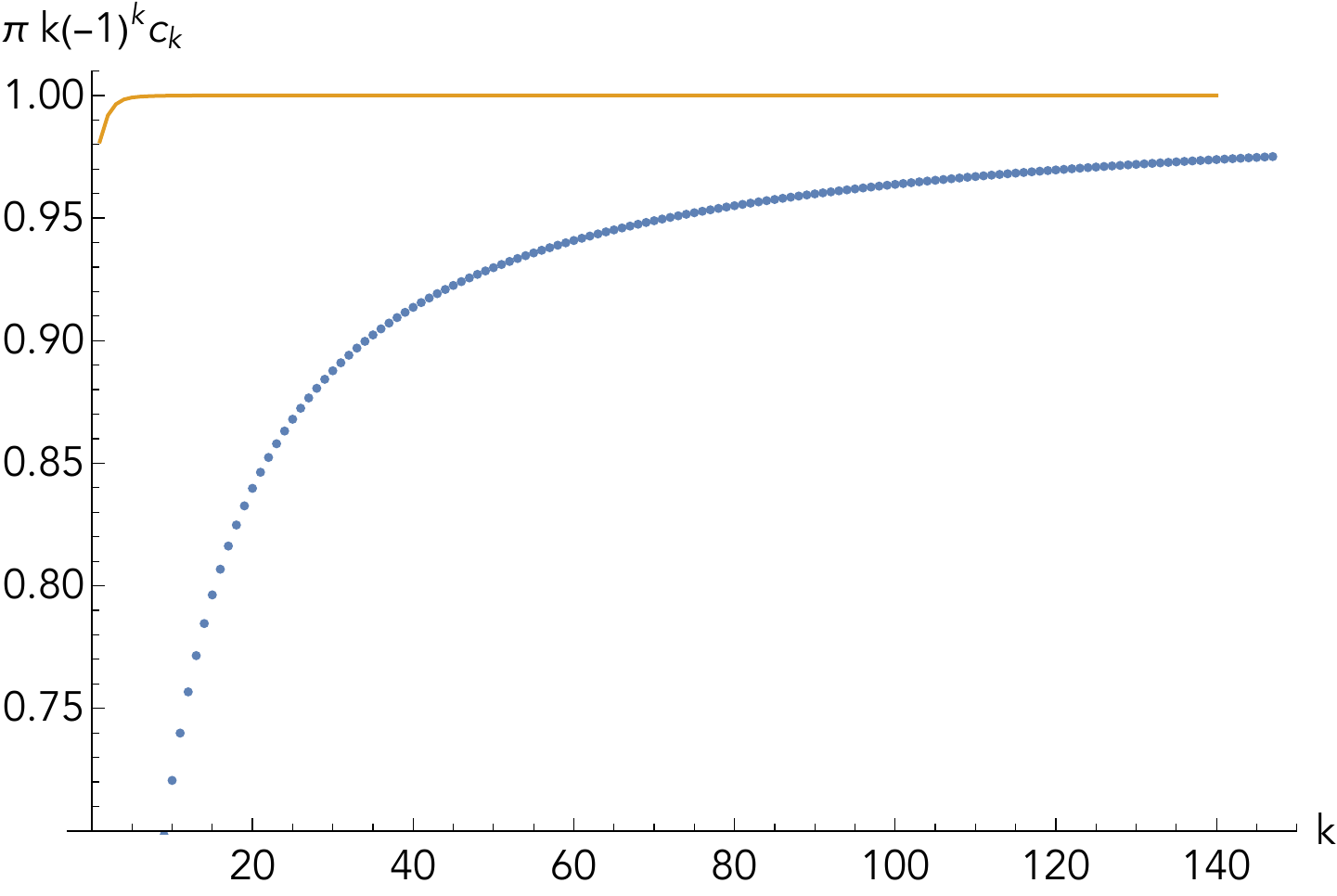}
\caption{\small $\tr M$ ratio test. On the left, we show in blue the ratio $c_{k+1}/c_{k}$ that tends to $-1$. The orange line is obtained after applying a 5th order Richardson acceleration. On the right, we present
the same analysis for the Darboux indicator $\pi k (-1)^{k}c_{k}$. }
\label{fig:trm1-ratio}
\end{center}
\end{figure}
%%%%%%%%%%%%%%%%%%%%%%%
%%%%%%%%%%%%%%%%%%%%%%%%%%%%%%%%
\begin{figure}[htb]
\begin{center}
\includegraphics[width=0.4\textwidth]{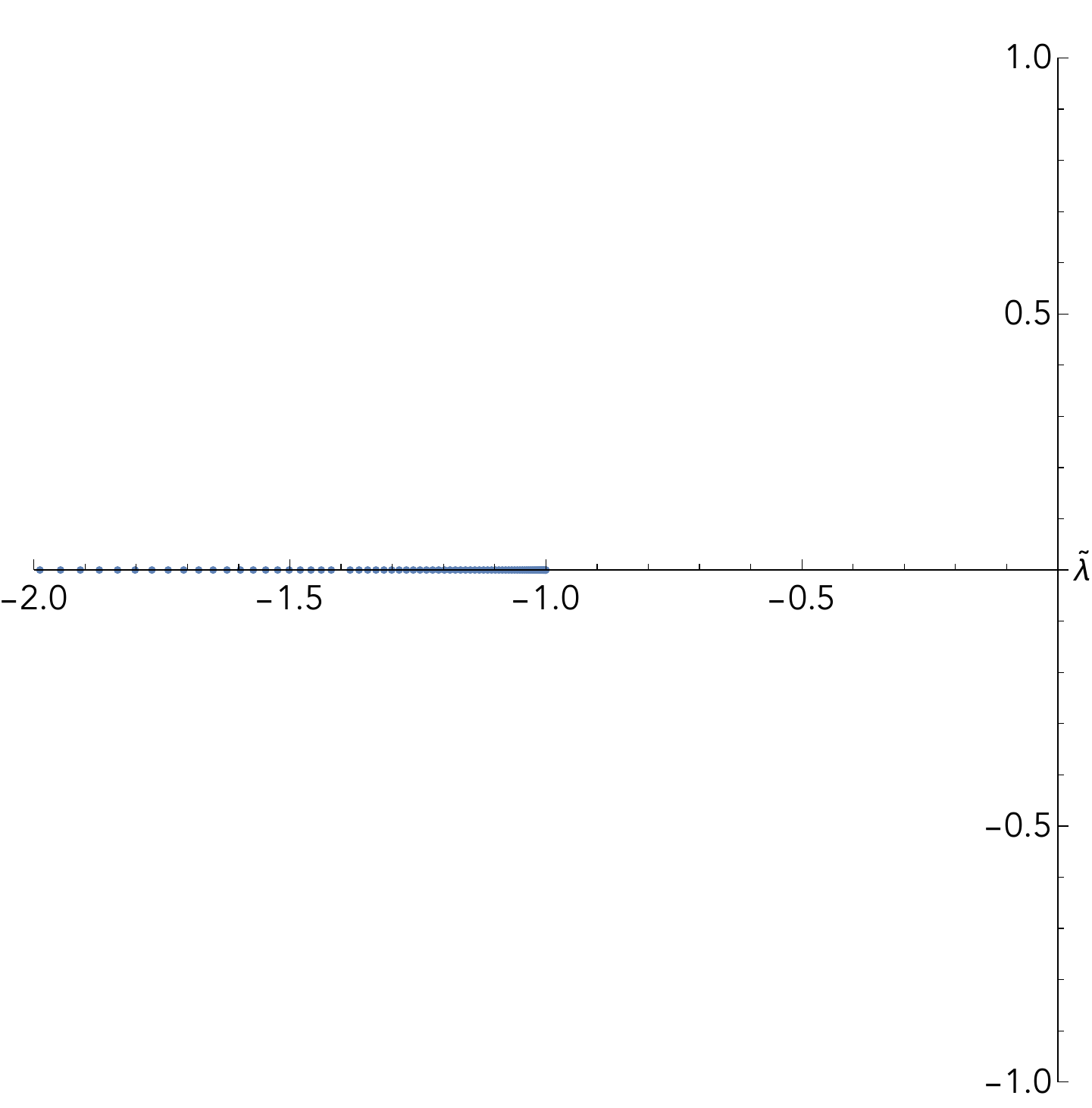} \hskip 0.15\textwidth
\includegraphics[width=0.4\textwidth]{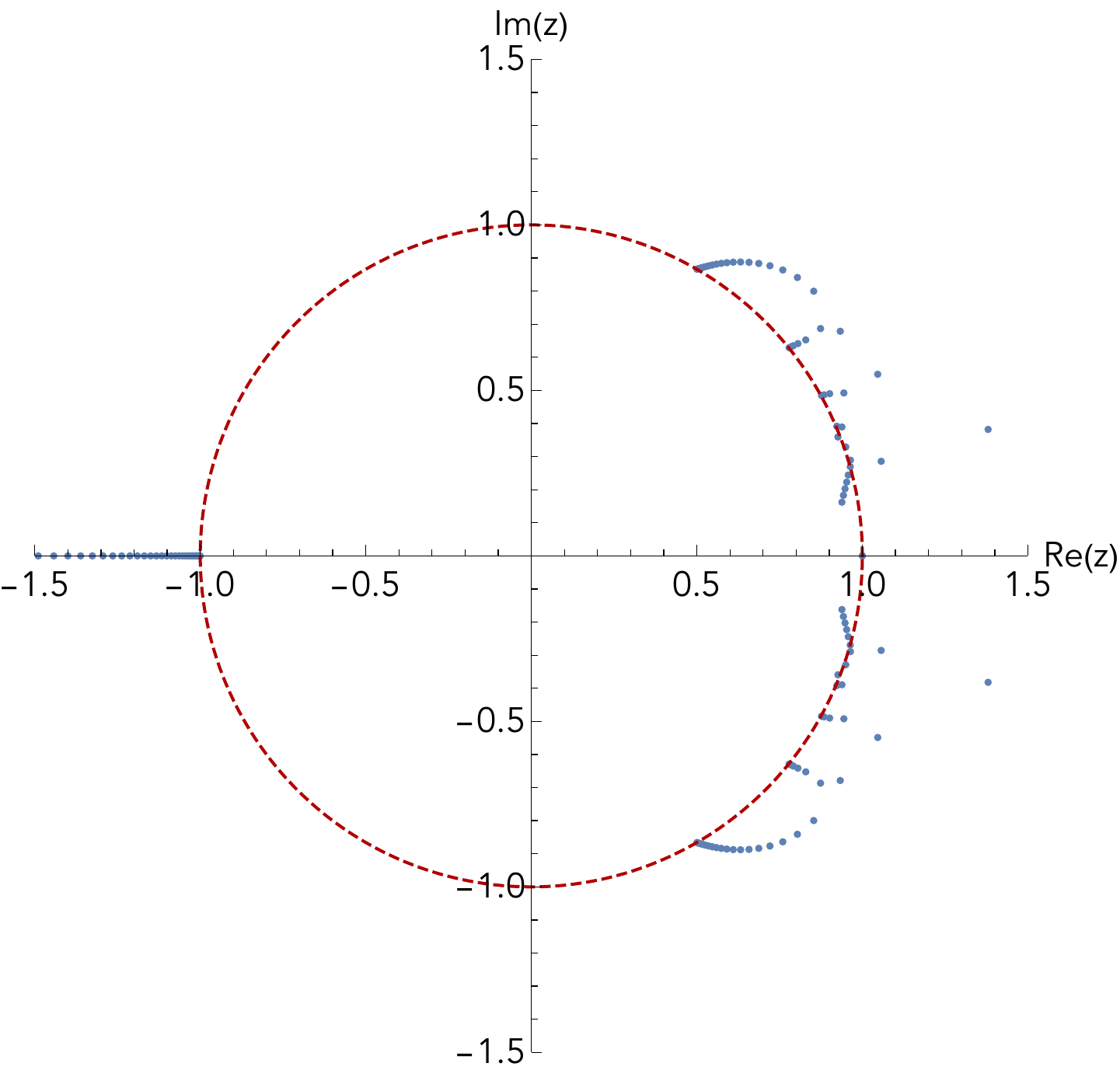}
\caption{\small Pad\'e poles of $\tr M$  from 150 terms. On the left, we show the poles of the direct approximants. These poles lie on the negative real axis and accumulate to $\tilde\lambda=-1$. On the right
we show the  poles in $z$-plane after application of the
conformal transformation (\ref{6.4}) followed by Pad\'e. In this case collinear singularities on the line $\tilde\lambda\in (-\infty, -1]$ are separated and made visible as arcs converging to points on the unit circle in the $z$ plane. These agree with a similar analysis for the 
whole $\Delta F$, see (\ref{6.5}).}
\label{fig:trm1-pade-poles}
\end{center}
\end{figure}
%%%%%%%%%%%%%%%%%%%%%%%%%%%
%%%%%%%%%%%%%%%%%%%%%%%%%%%%%%%%%%%%%%%%%%%%
\begin{figure}[htb]
\begin{center}
\includegraphics[width=0.45\textwidth]{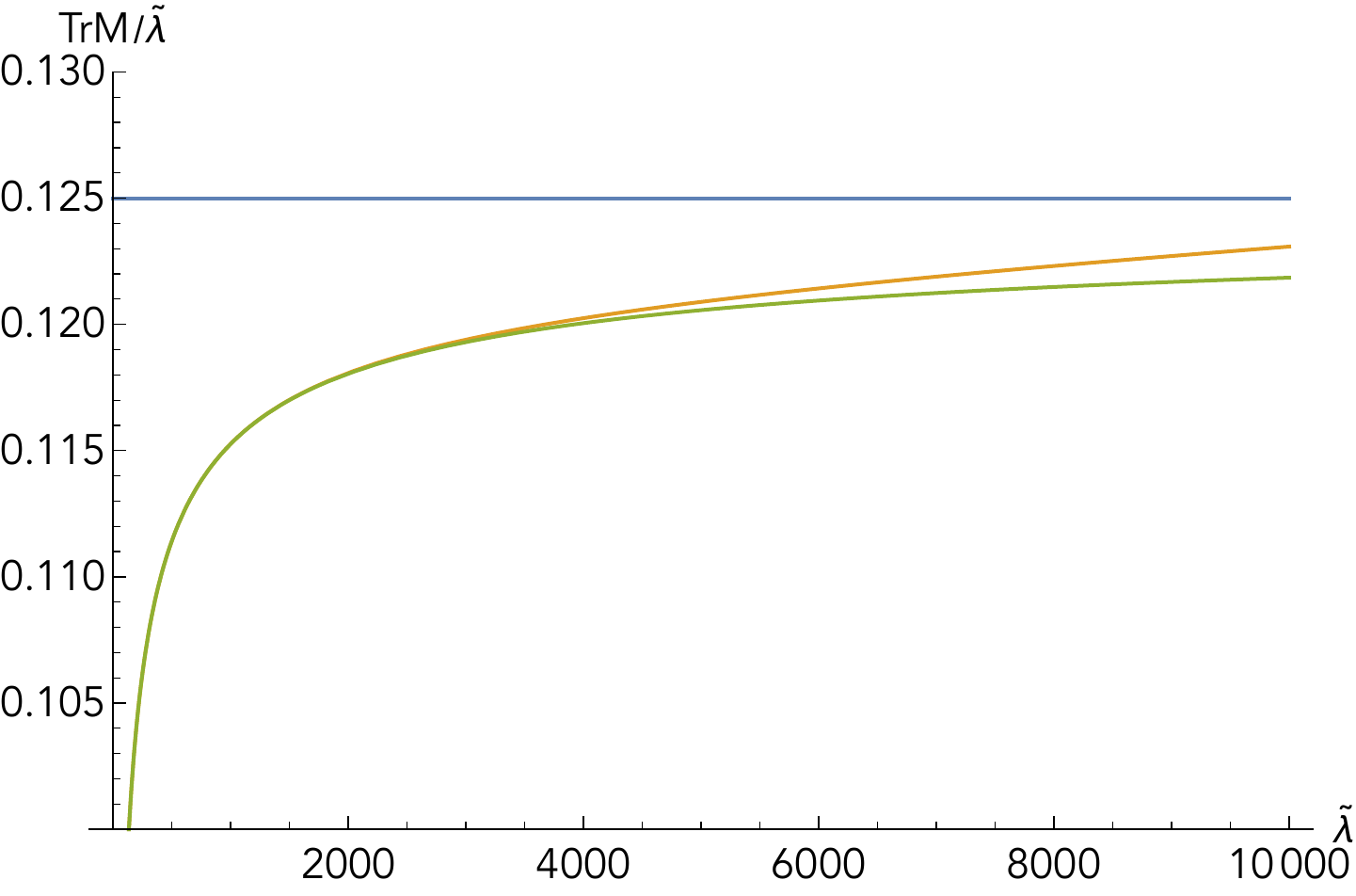} 
\includegraphics[width=0.45\textwidth]{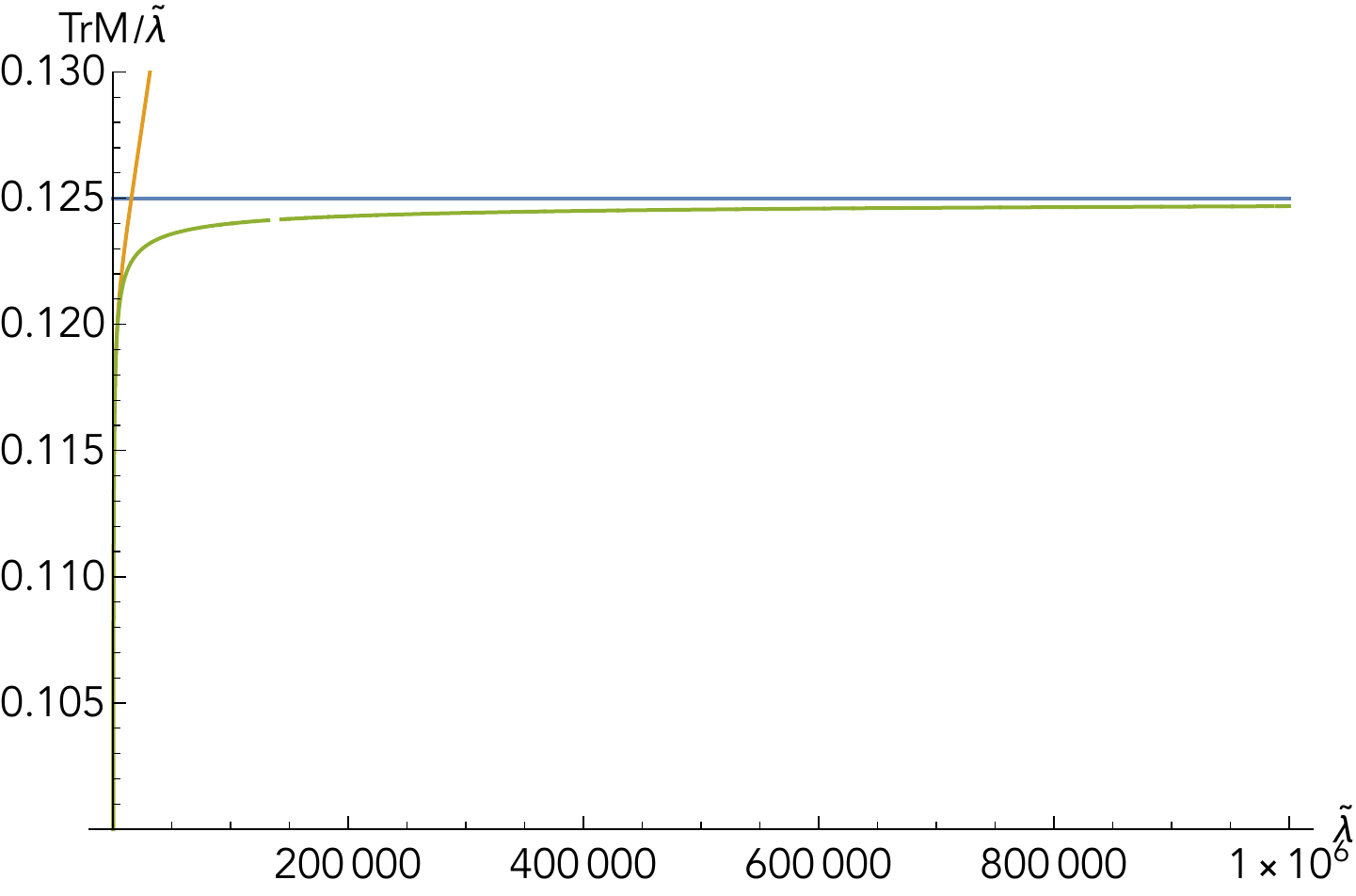}
\caption{\small Extrapolations of $\tr M(\tilde\lambda)/\tilde\lambda$ compared to the analytic value $1/8$ (blue line). The left and right plot differ only in the range of $\tilde\lambda$ values, $10^{4}$ on the left
and $10^{6}$ on the right. The orange line is diagonal Pad\'e of order 75, applied to the first 150 terms in the weak coupling expansion (\ref{6.6}). The
green line is the Pad\'e-conformal extrapolation based on the transformation (\ref{6.4}).}
\label{fig:trm1-extrap}
\end{center}
\end{figure}
%%%%%%%%%%%%%%%%%%%%%%%%%%%%%%%%%
%%%%%%%%%%%%%%%%%%%%%%%%%%%%%%%%%%%%%%%%%%%%%%%%%%%%%
\begin{figure}[h!]
\begin{center}
\includegraphics[width=0.45\textwidth]{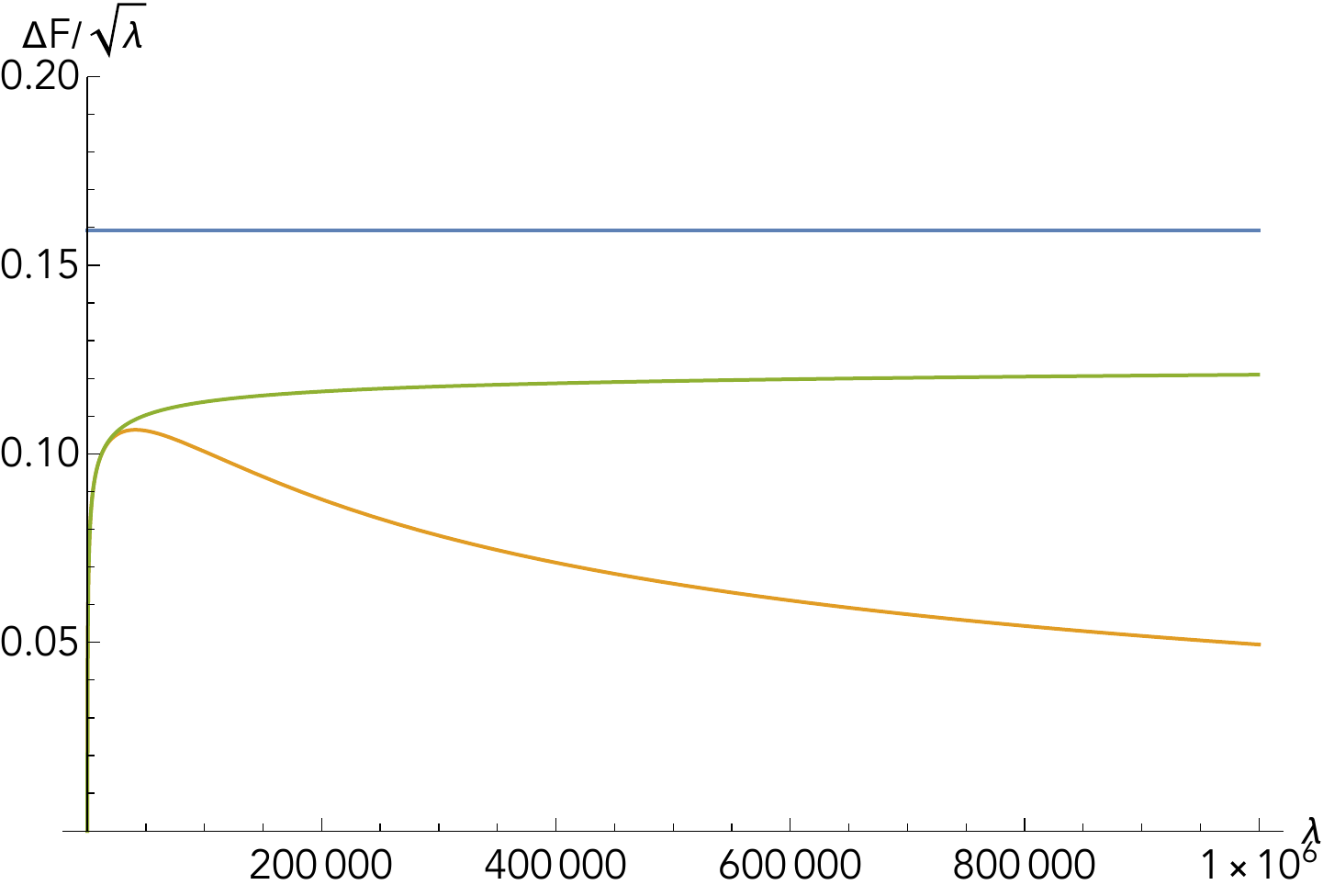}
\includegraphics[width=0.45\textwidth]{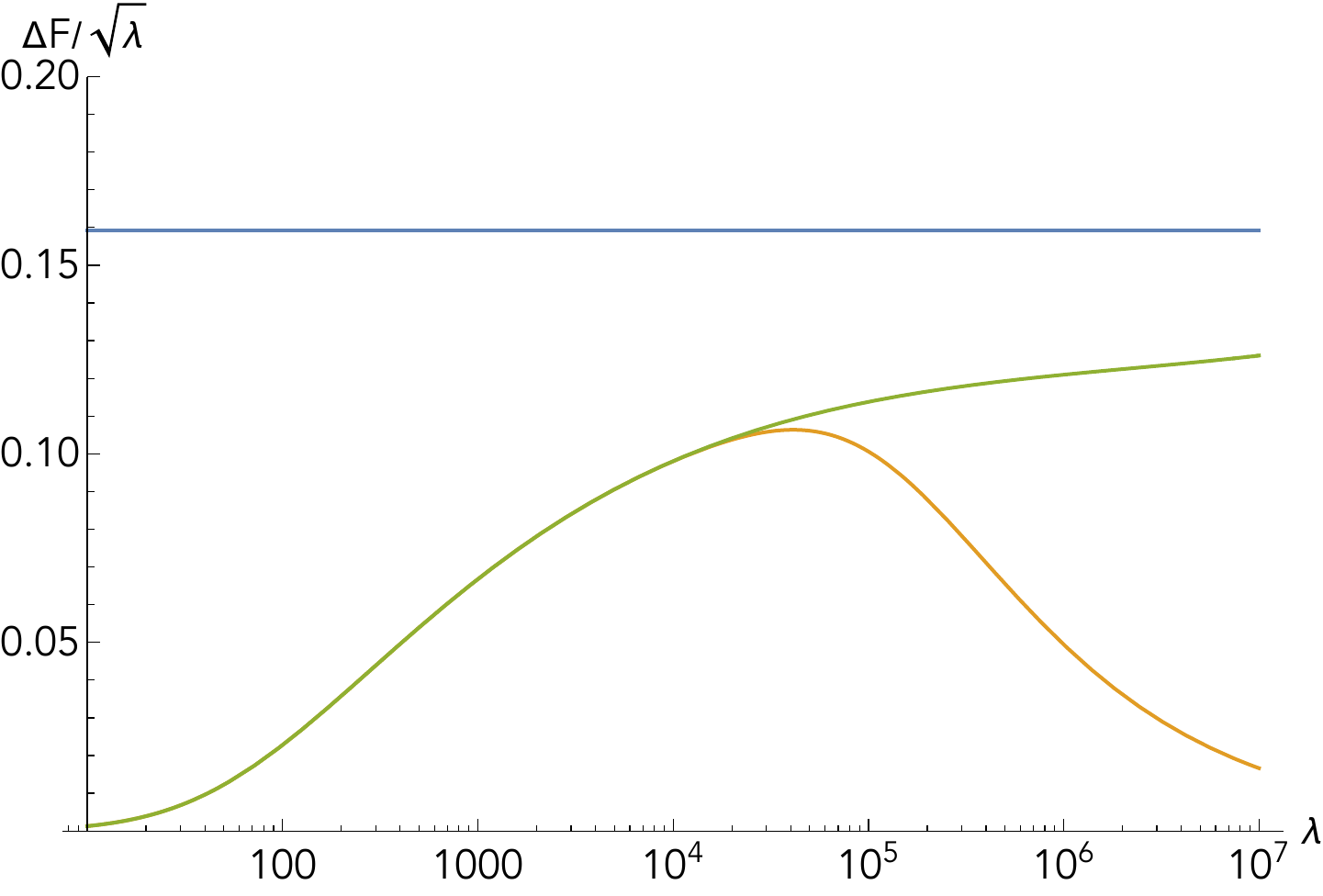}
\caption{\small Extrapolations of $\Delta F/\sqrt{\lambda}$ (plotted here as functions of $\l$,  not $\tilde\lambda$) compared to the asymptotic value $c_1= {1\over 2\pi}$ (blue line). In the left plot, in linear scale, the 
orange line is the diagonal Pad\'e approximant based on 150 terms of the full weak-coupling expansion, \ie the extension of (\ref{3.5}) to the order $\lambda^{150}$. This Pad\'e approximant breaks down shortly after $\lambda=10^{4}$. 
The green line is the Pad\'e conformal result and it extends to much higher values of the coupling $\lambda$. The left plot strongly supports the functional form $\Delta F \sim \sql$ at large $\lambda$. 
The convergence of the coefficient to the asymptotic value is steady but slow, as illustrated in the right panel on a logarithmic scale. See Section \ref{sec:deltaf} for a more refined estimate of the overall coefficient.}
%(in log scale) there is still a residual rising of the green curve and convergence to the asymptotic value is quite slow.}
\label{fig:trlog1pm-extrap}
\end{center}
\end{figure}
%%%%%%%%%%%%%%%%%%%%%%%%%%
The ratio $c_{k+1}/c_k$ is plotted in the left panel of Figure \ref{fig:trm1-ratio} based on the first 150 terms, indicating an alternating series with radius of convergence 1. 
The fact that the leading singularity is logarithmic is shown by the fact that $c_k\sim \frac{(-1)^k}{k} \times {\rm constant}$ as $k\to\infty$. See the right panel in Figure \ref{fig:trm1-ratio}.
%\begin{figure}[htb]
%\centerline{\includegraphics[scale=.6]{trm1-darboux-plot.pdf}}
%\caption{Blue = ratio. Gold = 5th order Richardson acceleration.}
%\label{fig:trm1-darboux}
%\end{figure}
The fact that the leading singularity is a branch point is also indicated by the Pad\'e poles, which are shown in the left panel of Figure \ref{fig:trm1-pade-poles}, accumulating along the negative real axis to the branch point at $\tilde\lambda=-1$.

After the conformal map (\ref{6.4}), followed by re-expansion to 150 terms in $z$, the poles of the resulting diagonal Pad\'e approximant are shown in the right panel of the same figure. 
This Figure indicates the existence of branch point singularities at the $z$ plane images of $\tilde\lambda=-1, -4,-9, -16$. The data becomes noisy at the conformal image of $-25$, 
with unphysical poles appearing inside the unit disk. These can be resolved by taking more terms in the original expansion.

We now map this Pad\'e approximant back to the physical $\tilde\lambda$ plane using the inverse conformal map in (\ref{6.4}), and plot to large $\lambda$. Figure \ref{fig:trm1-extrap} compares the diagonal Pad\'e extrapolation (orange curve), divided by $\tilde\lambda$, with the analytic large $\tilde\lambda$ limit of $\frac{1}{8}$ (blue curve) and the Pad\'e-conformal extrapolation (green curve). The first plot extends out to $\tilde\lambda=10^{4}$, while the second plot extends out to $\tilde\lambda=10^{6}$.
Note that the Pad\'e approximant eventually breaks down at $\tilde\lambda\approx 1.5\cdot 10^{4}$, while the Pad\'e-conformal approximant extends much further to very large $\tilde\lambda$. 
We stress that exactly the same input coefficient data was used in producing these two extrapolations, illustrating  the dramatic  effect of the conformal map.

 A similar analysis can be applied to 
$\tr M^{2}$ where we do not have a simple closed form expression for the expansion coefficients, but there is a systematic way to expand to very high order %, easily into the
 (multiple hundreds of terms, 
see (\ref{4.22})).
The resulting structure is very similar to that for the $\tr M$  case discussed  above, so we do not repeat the analogous plots.

\subsubsection{$\Delta F$}

Let us now   consider  the  large $\lambda$  extrapolation of  the full $\Delta F$. 
%To create a large $\lambda$ extrapolation of $\Delta F$,
 We begin with the small $\lambda$ expansion discussed in section \ref{sec:dF-weak}. 
We generated 150 terms of this expansion, with 450 digit precision for the coefficients. The coefficients are sums of products of  odd  $\zeta_{2k+1}$-values, but it is faster to work with finite but high precision coefficients.  
The ratio test and Pad\'e analysis again indicate a leading singularity at $\lambda=-\pi^2$, so we make the same conformal map (\ref{6.4}) and subsequent Pad\'e approximant 
and inverse map back to the physical $\lambda$ plane. 

Figure \ref{fig:trlog1pm-extrap} shows the result, and we again see that the Pad\'e-conformal extrapolation extends to a much larger value of $\lambda$.
This extrapolation shows that the functional form of the large $\lambda$ behavior is (left panel of the figure)
\be
\Delta F(\lambda)=\frac{1}{2} {\rm tr} \log(1+M) \stackrel{\lambda\to + \infty}{=} % {\rm constant}\times
c_1  \sqrt{\lambda} \ . 
%\quad \lambda\to + \infty.
\ee
This functional form matches the result of resumming the leading large $\lambda$ terms of $\tr M^{n}$ in \rf{5.10}, and the coefficients approximately agree. 

We stress that the only input information used for this extrapolation from small $\lambda$ to large $\lambda$ was the list of 150 perturbative coefficients. 
To get a better estimate of the result requires fitting the ratio $\Delta F/\sqrt\lambda$ and it is hard to support a specific functional form. The slow convergence shown in the right panel
of Figure \ref{fig:trlog1pm-extrap} should be
due to the expected logarithmic corrections in (\ref{1.14}) if they do not happen to cancel in $\Delta F$.

%%%%%%%%%%%%%%%%%%%%%%%%%%%%%%%%%%%%%%
\subsection{Evaluation of $\Delta F$ at  large $\lambda$ using truncation method}
\label{sec:deltaf}

In this subsection we use  a complementary numerical method in order to extract the precise large $\lambda$ behaviour of $\Delta F$. 
%
%
%
%We begin with the large $\lambda$ asymptotic expansion of the matrix elements $M_{ij}$ in (\ref{3.23}). Using again the Mellin transform
%methods we have  (\cf  (\ref{4.15}))
%\be
%M_{ij} = 8(-1)^{i+j}\sqrt{(2i+1)(2j+1)}\, N_{ij} \ ,  \la{68}
%\ee
%where 
%\be
%N_{ij} = \sql\,(f\star g_{ij})(\sql), \qquad \qquad 
%f(t) = \frac{e^{2\pi t}}{(e^{2\pi t}-1)^{2}},\qquad g_{ij}(t) = \frac{1}{t}J_{2i+1}(t)J_{2j+1}(t).
%\ee
%Evaluating the Mellin transforms and taking residues, we obtain the asymptotic expansion 
%\ba
%N_{ij}= & \Big[\frac{\delta_{ij}}{i(i+1)(2i+1)}+\frac{\delta_{i+1,j}}{(i+1)(2i+1)(2i+3)}+\frac{\delta_{i,j+1}}{i(2i-1)(2i+1)}\Big]\,\frac{\lambda}{64\pi^{2}}\notag\\
%& -\frac{\delta_{ij}}{24\,(2i+1)}+\frac{\zeta_3 }{2\pi^{2}}\,\cos\big(\pi(i-j)\big)\,  \frac{1}{\sql}+\cdots. \la{610}
%\ea
%At leading order in large $\lambda$ we keep only the term linear in $\lambda$ in $M$ in \rf{68},\rf{610}. 
%As a check, we can recover the  $\tr M^n$ asymptotics  (\ref{5.1}),(\ref{5.5}). For example, 
%\ba
%\tr M &= \sum_{i=1}^{\infty}8(2i+1)\Big[\frac{1}{64\pi^{2}\,i(i+1)(2i+1)}\,\lambda+\cdots\Big] = \frac{\lambda}{8\pi^{2}}+\cdots, \notag \\
%\tr M^{2} &= \frac{\lambda^{2}}{240\pi^{4}}+\frac{3\lambda^{2}}{32\pi^{4}}\sum_{i=2}^{\infty}\frac{1}{i(i+1)(2i-1)(2i+3)} +\cdots
%= \frac{\lambda^{2}}{192\pi^{4}}+\cdots \ . 
%\ea
%in agreement with our previous results. One checks similarly agreement of higher traces with previous results. 
Starting from the expansion (\ref{610}), 
let us denote by $M_K$  the $K\times K$ matrix  which is 
the linear in $\lambda$ part of $M$, truncated to the first $K$ rows and columns. Then 
\be
\la{6.12} \Delta F(\lambda) = \lim_{K \to \infty} \Delta F_K(\lambda) \ , \qquad \qquad 
\Delta F_K(\lambda) = \frac{1}{2}\tr\log(1+M_K).
\ee
To determine the large $\lambda$ behaviour of $\Delta F(\lambda)$, we need to take first $K\to \infty$, 
 and then  $\lambda\to \infty$.
 
\begin{figure}
\begin{center}
\includegraphics[width=0.45\textwidth]{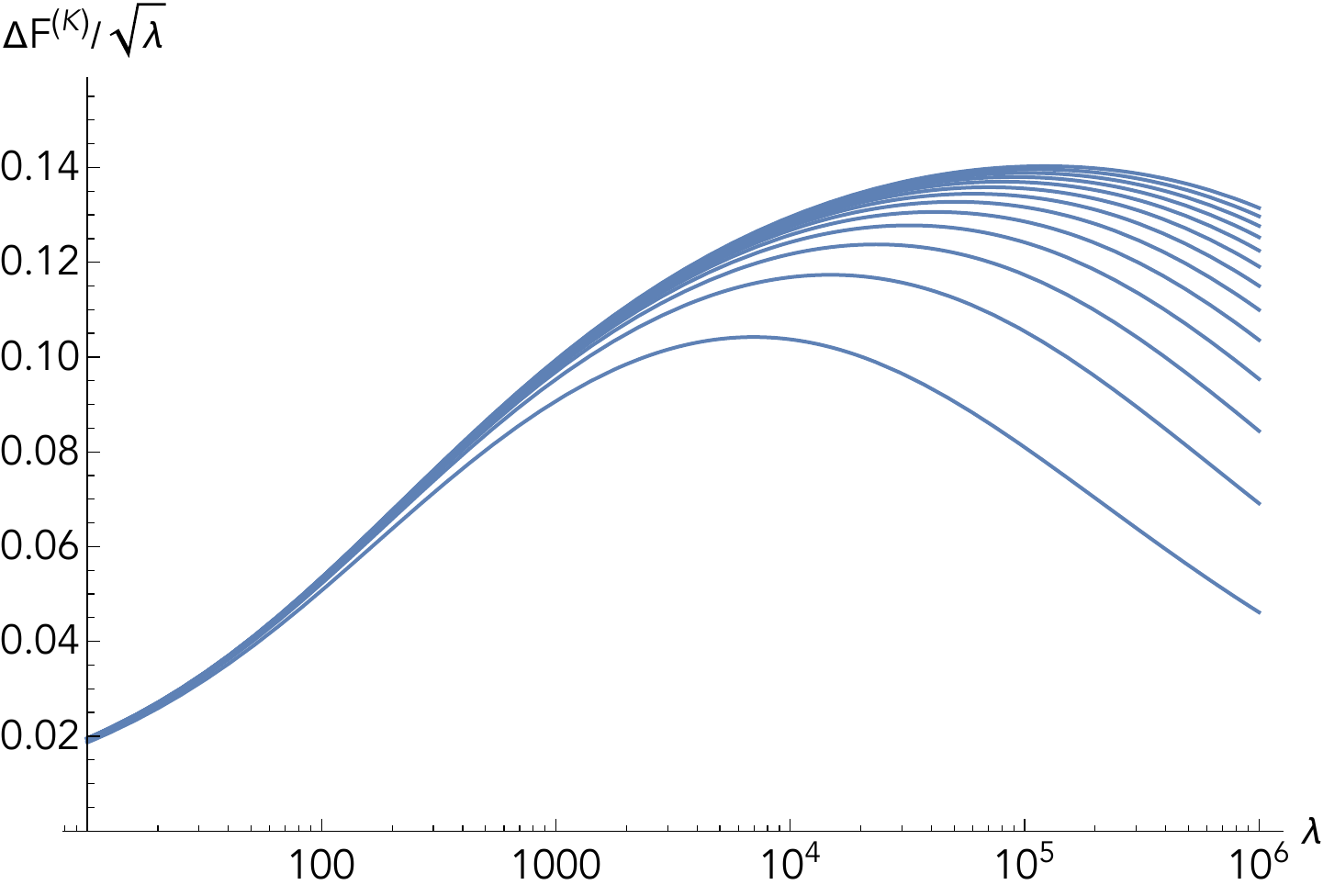}
\includegraphics[width=0.45\textwidth]{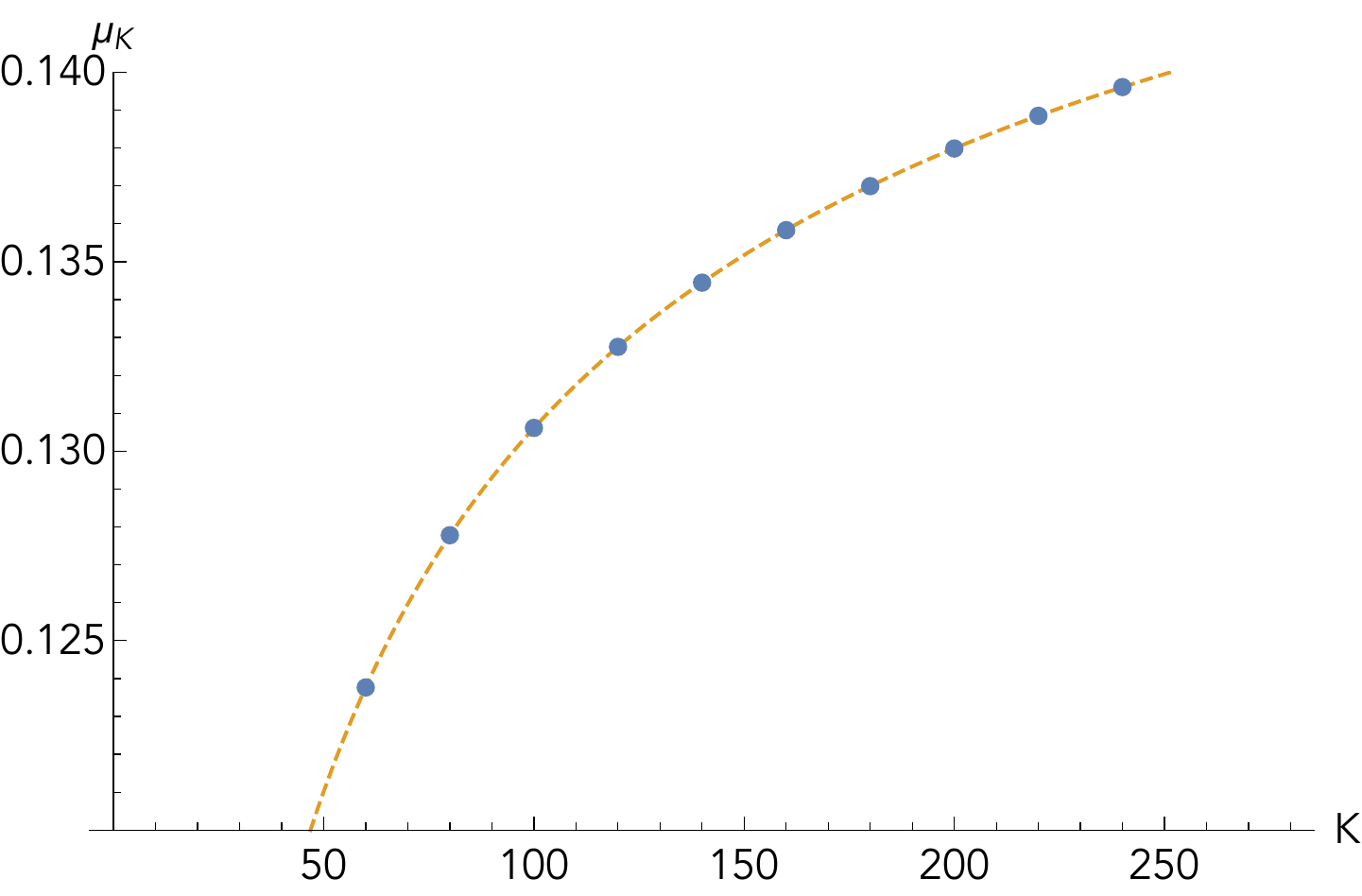}
\caption{\small Analysis of $\Delta F$ by considering a truncated leading order approximation of the matrix $M$. In the left panel 
we plot the ratio $\Delta F_K/\sql$ where $F_K$ is defined in (\ref{6.12}), and $K=20, 40, \dots, 260$ from bottom to top.
For each $K$, there is a maximal  value $\mu_{K}$. In the right panel we plot $\mu_{K}$ vs. $K$ (blue dots) and compute its best fit
(orange dashed line) with a constant plus a leading  $\sim K^{-1/2}$ and  subleading  $\sim K^{-1}$ terms. The best fit
parameters are in (\ref{6.14}).
}
\label{fig:trunc}
\end{center}
\end{figure}

To bypass this double limit procedure, we will fix $K$, increase $\lambda$ until the ratio $\frac{\Delta F_K(\lambda)}{\sql}$ reaches a maximum
\be
\la{6.13}
\mu_{K} = \max_{\lambda}\frac{\Delta F_K(\lambda)}{\sql}\ ,
\ee
and,  finally,  extrapolate $\mu_{K}$ to $K\to \infty$.
 According to (\ref{5.11}), the expected value is $c_1=\frac{1}{2\pi}$. The explicit numerical 
results are collected in Figure \ref{fig:trunc}. In the left panel we show the curves $\frac{\Delta F_K(\lambda)}{\sql}$ for $K=20,40,60, \dots 260$.
For each $K$ a maximum in \rf{6.13} is reached at a value of $\lambda$ that increases with $K$. The maximum value
 $\mu_{K}$ is shown in the right panel
of the figure and fitted by the dashed curve
\be
\la{6.14}
\mu_{K}^{\rm fit} = 0.158-\frac{0.301}{\sqrt K}+\frac{0.290}{K},
\ee
that empirically works very well. The estimated value of the coefficient of $\sql$ in (\ref{5.11}) is thus $0.158$, which  differs by less then 1\%
from the analytical prediction $\frac{1}{2\pi} \simeq 0.159$.

\subsection*{Acknowledgements}
We would like to thank M. Bill\`o, S. Giombi, A. Lerda  and  J. Russo   for related discussions. 
MB was supported by the INFN grant GSS (Gauge Theories, Strings and Supergravity). 
GD was supported by the U.S. Department of Energy, Office of Science,
Office of High Energy Physics under Award Number DE-SC0010339.
AAT was supported by the STFC grant ST/T000791/1.

\appendix

%\newpage
\section{Direct  computation  of $\Delta q$   at weak coupling \la{app-Dq}}

The expectation value of the Wilson loop \rf{2.7}    at finite $N$ reads 
\be
%v3
\vev{\mc W}^{\orientif} = N \Big[ \vev{\mc W}'_{0}+\frac{1}{N^{2}}\,\Big(\vev{\mc W}_{1}^{\mc N=4}+\vev{\mc W}_{1}^{\N=2} \Big)+\mc O\Big({1\ov N^{4}}\Big)\Big],
\ee
where $N \vev{\mc W}'_{0} =\vev{\mc W}_{0}^{\mc N=4}$ is the planar  $\N=4$ SYM expression   (\ref{1.1}) %(here we add an explicit $\mc N=4$ label for clarity)
 and $\vev{\mc W}_{1}^{\mc N=4} $ is  \cite{Drukker:2000rr} (cf. \rf{1.3})
\be
\vev{\mc W}_{1}^{\mc N=4} = \tfrac{1}{48}\,\Big[-12\sql I_{1}(\sql)+\lambda I_{2}(\sql)\Big] \ .
\ee
We use the label ``$\N=2$'' to separate the genuine   correction to the $\N=4$ result. 
The   explicit calculation starting with the matrix model expression \rf{2.7},\rf{3.3}
gives % the genuine $\mc O(1/N^{2})$ correction in the $\mc N=2$ orientifold theory
\ba
\la{H.3}
\te  \vev{\mc W}_{1}^{\N=2} = &\te -\frac{15 \zeta_{5}}{4\,(8\pi^{2})^{3}}\,\lambda^{4}
+\Big(-\frac{15 \zeta_{5}}{32\,(8\,\pi^{2})^{3}}+\frac{105 \zeta_{7}}{2\,(8\,\pi^{2})^{4}}\Big)\,\lambda^{5}
+\Big(-\frac{5 \zeta _5}{256\,(8\pi^{2})^{3}}+\frac{105 \zeta _7}{16\,(8\pi^{2})^{4}}-\frac{2205 \zeta _9}{4\,(8\pi^{2})^{5}}\Big)\,\lambda^{6}  \lp \te
+\Big(-\frac{5 \zeta _5}{12288\,(8\pi^{2})^{3}}+\frac{75 \zeta _5^2}{2\,(8\pi^{2})^{6}}+\frac{35 \zeta _7}{128\,(8\pi^{2})^{4}}
-\frac{2205 \zeta _9}{32\,(8\pi^{2})^{5}}
+\frac{10395 \zeta _{11}}{2\,(8\pi^{2})^{6}}\Big)\,\lambda^{7}+\mc O(\lambda^{8}).
\ea
The function $\Delta q(\l) $ in \rf{1.2},\rf{1.10}   is obtained dividing by $\vev{\mc W}_{0}$ and this gives precisely (\ref{3.7}), \ie the result  consistent with  (\ref{3.6}). 
We checked the  relation (\ref{3.6})  to  order $\mc O(\lambda^{20})$ by an independent computation of both $\Delta F$ and $\Delta q$.

Let us note that 
 each of the monomials in the $\zeta_n$-values appears  in the weak-coupling expansion 
 with a simple single-power  dependence on $\lambda$. 
% Consistently, 
 The corresponding leading $1/N^{2}$ corrections in $\vev{\mc W}^{\orientif}$  happen to  have its non-trivial dependence on $\l$ 
  via the Bessel function  factor  $\sim\l^{-1/2}  I_{1}(\sql)$.
 This   property can be proved for specific monomials in $\zeta_n$-values by the methods described in \cite{Beccaria:2021ksw}.
It may be  made explicit  by collecting terms in (\ref{H.3})  as %according to 
\ba
 \vev{\mc W}_{1}^{\mc N=2} =&\te  -\frac{15 \zeta_{5}}{4\,(8\pi^{2})^{3}}\,\lambda^{4}\,\Big(
1+\frac{\lambda}{8}+\frac{\lambda^{2}}{192}+\frac{\lambda^{3}}{9216}+\cdots\Big) 
%-\frac{15 \zeta_{5}}{32\,(8\,\pi^{2})^{3}}\lambda^{5}-\frac{5 \zeta _5}{256\,(8\pi^{2})^{3}}\lambda^{6}-\frac{5 \zeta _5}{12288\,(8\,\pi^{2})^{3}}\lambda^{7}
%+\cdots \lp
%%%
\lp \te +\frac{105 \zeta_{7}}{2\,(8\pi^{2})^{4}}\,\lambda^{5}\,\Big(
1+\frac{\lambda}{8}+\frac{\lambda^{2}}{192}+\cdots
\Big)
%+\frac{105 \zeta _7}{16\,(8\pi^{2})^{4}}\lambda^{6}
%+\frac{35 \zeta _7}{128\,(8\pi^{2})^{4}}\lambda^{7}+\cdots \lp
%%%
 -\frac{2205 \zeta _9}{4\,(8\pi^{2})^{5}}\,\lambda^{6}\,\Big(1+\frac{\lambda}{8}+\cdots\Big) + \cdots.
 \ea
 Here, one can see that each monomial in the $\zeta_n$-values is multiplied by  the   %a round bracket which is in all cases the 
 expansion 
 of  the factor  $\vev{\mc W}_{0}=\frac{2}{\sql}\,I_{1}(\sql)$. 
 As discussed in \cite{Beccaria:2021ksw}, this property is important for the  relation (\ref{3.6})  to hold. 

\section{Weak coupling expansion of $\tr M^{2}$}
\la{app:M2}

Here   we shall provide the proof of  the result (\ref{4.22}) for the weak-coupling expansion of $\tr M^{2}$. 
We begin with the expansion of the product of two Bessel functions as a series of Jacobi polynomials
\be
2 t t' J_{2}(t\sqrt u)J_{2}(t'\sqrt u) =2\,(tt')^{3}\, \sum_{m=0}^{\infty}(-1)^{m} \frac{u^{m+2}}{4^{m+2}\big[\Gamma(m+3)\big]^{2}} t^{2m} P_{m}^{(2,-2m-5)}\Big(1-2\,\frac{t'^{2}}{t^{2}}\Big).
\ee
From (\ref{4.7}), the kernel in (\ref{4.3}) admits the representation 
\be
G(t\sql,t'\sql) = 2\,(tt')^{3}\,\sum_{m=0}^{\infty}\lambda^{m+3} \frac{(-1)^{m}}{4^{m+2}(m+3)\big[\Gamma(m+3)\big]^{2}} t^{2m} P_{m}^{(2,-2m-5)}\Big(1-2\,\frac{t'^{2}}{t^{2}}\Big).
\ee
Plugging this into (\ref{4.5}) gives
\be
\la{A.3}
\tr M^{2} = 8\sum_{\ell=0}^{\infty}\lambda^{\ell+6}\int_{0}^{1} dx x^{5}d_{\ell}(x)\int_{0}^{\infty}dt\frac{e^{2\pi t}}{(e^{2\pi t}-1)^{2}}\frac{e^{2\pi x t}}{(e^{2\pi x t}-1)^{2}}\,t^{2\ell+11},
\ee
where the polynomials $d_{\ell}(x)$ were defined in (\ref{4.20}) and (\ref{4.21}).
Using $d_{\ell}(1/x) = x^{-2\ell}d_{\ell}(x)$, we get 
\ba
\la{A.4}
&\tr M^{2} = 8\sum_{\ell=0}^{\infty}f_{\ell}\,\lambda^{\ell+6},
\lp
 f_{\ell} = \int_{0}^{1} dx x^{5}d_{\ell}(x)\int_{0}^{\infty}dt\frac{e^{2\pi t}}{(e^{2\pi t}-1)^{2}}\frac{e^{2\pi x t}}{(e^{2\pi x t}-1)^{2}}\,t^{2\ell+11} \lp
%=\int_{1}^{\infty} \frac{dx}{x^{2}} x^{-5}d_{\ell}(1/x)\int_{0}^{\infty}dt\frac{e^{2\pi t}}{(e^{2\pi t}-1)^{2}}\frac{e^{2\pi  t/x}}{(e^{2\pi  t/x}-1)^{2}}\,t^{2\ell+11} \lp
=\int_{1}^{\infty} \frac{dx}{x^{2}} x^{-5}x^{-2\ell}d_{\ell}(x)\int_{0}^{\infty}dt\frac{e^{2\pi t}}{(e^{2\pi t}-1)^{2}}\frac{e^{2\pi  t/x}}{(e^{2\pi  t/x}-1)^{2}}\,t^{2\ell+11} \lp
%=\int_{1}^{\infty} dx  x^{-2\ell-7}d_{\ell}(x)\int_{0}^{\infty}dt\,x\,\frac{e^{2\pi x t}}{(e^{2\pi x t}-1)^{2}}\frac{e^{2\pi  t}}{(e^{2\pi  t}-1)^{2}}\,(tx)^{2\ell+11} 
=\int_{1}^{\infty} dx  x^{5}d_{\ell}(x)\int_{0}^{\infty}dt\,\frac{e^{2\pi x t}}{(e^{2\pi x t}-1)^{2}}\frac{e^{2\pi  t}}{(e^{2\pi  t}-1)^{2}}\,t^{2\ell+11}.
\ea
 $f_{\ell}$ may be written as an integral over the whole half-line $[0,\infty]$ and have (cf. \rf{4.21})
\ba
f_{\ell} &=\frac{1}{2} \int_{0}^{\infty} dx x^{5}d_{\ell}(x)\int_{0}^{\infty}dt\frac{e^{2\pi t}}{(e^{2\pi t}-1)^{2}}\frac{e^{2\pi x t}}{(e^{2\pi x t}-1)^{2}}\,t^{2\ell+11} \lp
%= \frac{1}{2} \int_{0}^{\infty} dx x^{5}d_{\ell}(x)\int_{0}^{\infty}dt\frac{e^{2\pi t}}{(e^{2\pi t}-1)^{2}}\frac{e^{2\pi x t}}{(e^{2\pi x t}-1)^{2}}\,t^{2\ell+11} \lp
= \sum_{m}a_{m}^{(\ell)}\frac{1}{2} \int_{0}^{\infty} dx x^{5+\ell}(x^{m}+x^{-m})\int_{0}^{\infty}dt\frac{e^{2\pi t}}{(e^{2\pi t}-1)^{2}}\frac{e^{2\pi x t}}{(e^{2\pi x t}-1)^{2}}\,t^{2\ell+11}\lp
=  \frac{1}{2}\,\sum_{m}a_{m}^{(\ell)} (I_{\ell+m+5}I_{\ell+5-m}+I_{\ell-m+5}I_{\ell+5+m}) = \sum_{m}a_{m}^{(\ell)} I_{\ell+m+5}I_{\ell+5-m},
\ea
where
\be
I_{n} = \int_{0}^{\infty}dt \frac{e^{2\pi t}}{(e^{2\pi t}-1)^{2}}t^{n} = (2\pi)^{-n-1}\Gamma(n+1)\zeta_n \ .
\ee
Hence, 
\be
f_{\ell} = (2\pi)^{-12-2\ell}\sum_{m}a_{m}^{(\ell)} \Gamma(\ell+6+m)\Gamma(\ell+6-m)\, \zeta_{\ell+5+m}\, \zeta_{\ell+5-m}\ .
\ee
Combined with \rf{A.4},  this  proves the relation  (\ref{4.22}).

\section{Coefficients $b_n$ in  strong coupling limit of $\tr M^n$}
\la{NEW}

Here we shall  discuss  the explicit form of the  coefficients $b_n$ in \rf{5.1}. They can be computed 
by explicit evaluation of the traces $\tr {\rm S}^{n}$ of $\S$ in  \rf{59}, since the infinite sums $\sum_{i=1}^{\infty} \S_{ii}$, 
$\sum_{i,j=1}^{\infty}\S_{ij}\S_{ji}$, etc. are all convergent.  For instance, 
\ba
b_{1} = \tr \S = \sum_{i=1}^{\infty}\tfrac{1}{4i(i+1)} = \tfrac{1}{4}, \qquad
b_{2} = \tr \S^{2} = \tfrac{1}{60}+\frac{3}{8}\sum_{i=2}^{\infty}\tfrac{1}{i(i+1)(2i-1)(2i+3)} = \tfrac{1}{48}, \ \dots\ .
\ea
An alternative representation for $b_n$  that avoids  infinite summations is found using 
%is based on 
%Note  that $b_n$ are   related  to $C_n$ in  \rf{4.2} as $C_n = {(-1)^{n+1}\ov n (2 \pi^2)^n}    b_n $. 
%Using that  
$\frac{1}{t}\frac{e^{2\pi t/\sql}}{(e^{2\pi t/\sql}-1)^{2}} = \frac{\lambda}{4\pi^{2}t^{3}}-\frac{1}{12\,t}+\cdots$
%\be
%\frac{1}{t}\frac{e^{2\pi t/\sql}}{(e^{2\pi t/\sql}-1)^{2}} = \frac{\lambda}{4\pi^{2}t^{3}}-\frac{1}{12\,t}+\cdots
%\ee
and the integral representation (\ref{4.7}):
\ba
\la{5.3}
&b_{n} = \int_{0}^{1} dx_{1}\cdots \int_{0}^{1}dx_{n}\ f(x_{1}, x_{2})\,f(x_{2},x_{3})\cdots f(x_{n-1}, x_{n})\, f(x_{n}, x_{1}),
\\
\la{5.4}
& f(x,y) = \int_{0}^{\infty}\frac{dt}{t}J_{2}(t\,\sqrt x)\,J_{2}(t\,\sqrt y) = \begin{cases} 
\frac{y}{4x}, & x\ge y, \\
\frac{x}{4y}, & x< y\ .
\end{cases}
\ea
The explicit values of the coefficients $b_{n}$ can be  easily computed from \rf{5.3} %and one finds
\ba
\la{5.5}
&\{b_{n}\}_{n=1,2, \dots} = \big\{
\tfrac{1}{4},\tfrac{1}{48},\tfrac{1}{384},\tfrac{1}{2880},\tfrac{13}{
276480},\tfrac{11}{1720320},\tfrac{647}{743178240},\tfrac{1133}{
9555148800},\tfrac{43213}{2675441664000}, \dots
%\tfrac{777013}{353158299648000},   \lp
%\tfrac{2540291}{8475799191552000},\tfrac{31489807}{771297726431232000},
%\tfrac{14413012699}{2591560360808939520000},
%\tfrac{471206866243}{
%621974486594145484800000},\dots
\big\}.
\ea
A much more efficient way to determine $b_n$  is based on using  (\ref{555}) since it implies
the following explicit expression for their generating function
\be
\la{C.5}
b(x) = \sum_{n=1}^{\infty} b_{n}x^{n-1} = \frac{1}{\sqrt{2x}}\frac{J_{2}(\sqrt{2x})}{J_{1}(\sqrt{2x})}.
\ee
Expanding (\ref{C.5}) at small $x$  %immediately
 reproduces the values (\ref{5.5}).
 
To prove \rf{C.5} let us   first  note  that (\ref{5.10}) implies that 
\be
\la{C.6}
b_{n} = \tr {\rm S}^{n} = 2^{n}\,\sigma_{n},\qquad\qquad  \sigma_{n} = \sum_{k=1}^{\infty}\frac{1}{{\rm j}_{1,k}^{2n}}\ ,
\ee
where  $\{\jj_{1,k}\}$ are  the (positive) zeroes  of the Bessel   function $J_{1}(x)$.
The generating function (\ref{C.5}) may then be obtained as a corollary of the results in the recent paper \cite{ciaurri2018}  that 
 proved that 
%Specializing their results to the case of $J_{1}$ and defining
%\be
%\sigma_{n} = \sum_{k=1}^{\infty}\frac{1}{{\rm j}_{1,k}^{2n}}\ ,
%\ee
%one has 
\be
\sigma_{n} = \frac{(-1)^{n+1}}{2^{2n}n!(2)_{n-1}}\mathsf{B}_{2n,0}(1)\ ,
\ee
where $\mathsf{B}_{2n,0}(x)$ are a special case of the Bernoulli-Dunkl polynomials. They  are generated  by 
%\be
%\frac{1}{\mc I_{1}(t)}E_{0}(xt) = \sum_{n=0}^{\infty}\mathsf{B}_{n,0}(x)\frac{t^{n}}{\gamma_{n,0}}, \qquad 
%\gamma_{n,0} = \begin{cases}
%2^{2k}k!(1)_{k}, & n=2k, \\
%2^{2k+1}k!(1)_{k+1}& n=2k+1.
%\end{cases}
%\ee
%\be
%{\mc I}_{0}(z) = J_{0}(iz), \qquad {\mc I}_{1}(z) = 2\frac{J_{1}(iz)}{iz},
%\qquad 
%E_{0}(z) = {\mc I}_{0}(z)+\frac{z}{2}{\mc I}_{1}(z) = I_{0}(z)+I_{1}(z).
%\ee
%\be
%\frac{t}{2}+\frac{t}{2}\frac{I_{0}(t)}{I_{1}(t)} = \sum_{n=0}^{\infty}\mathsf{B}_{n,0}(1)\frac{t^{n}}{\gamma_{n,0}}
%\ee
\be\la{c8}
\frac{t}{2}+\frac{t}{2}\frac{I_{0}(t)}{I_{1}(t)} -1 = \sum_{m=1}^{\infty}\mathsf{B}_{m,0}(1)\frac{t^{m}}{\gamma_{n,0}}, \qquad \qquad \gamma_{m,0} = 
\begin{cases}
2^{2n}\, k!\,(1)_{n}, & \ \ m=2n, \\
2^{2n+1}\, n!\,(1)_{n+1}, & \ \ m=2n+1.
\end{cases}
\ee
Taking the even in $t$  part of \rf{c8}  gives 
\ba
& \frac{t}{2}\frac{I_{0}(t)}{I_{1}(t)} -1 = \sum_{n=1}^{\infty}\mathsf{B}_{2n,0}(1)\frac{t^{2n}}{\gamma_{2n,0}}  = \sum_{n=1}^{\infty} (-1)^{n+1}\sigma_{n}t^{2n}.
\ea
Finally, comparing with (\ref{C.5}) and (\ref{C.6}), we  get the proof  of \rf{C.5}
\be
b(x) = \sum_{n=1}^{\infty}\, 2^{n} \, \sigma_{n}\, x^{n-1} = -\frac{1}{x}\bigg[i\sqrt\frac{x}{2}\,\frac{I_{0}(i\,\sqrt{2x})}{I_{1}(i\,\sqrt{2x})} -1\bigg]
= \frac{1}{\sqrt{2x}}\frac{J_{2}(\sqrt{2x})}{J_{1}(\sqrt{2x})} \ . 
\ee
%in agreement with (\ref{5.5}).
 By the same methods, the results in \cite{ciaurri2018}   can be   used to construct the generating function
for the sums $\sum_{k=1}^{\infty}\frac{1}{{\rm j}_{a,k}^{2n}}$ of inverse negative even powers of zeroes of $J_{a}(x)$.

\section{Darboux theorem}
\la{app:darboux}

Darboux's theorem states that for a convergent series expansion, the large-order growth of the expansion coefficients about a point (say $t=0$) is directly related to the behaviour of the expansion in the vicinity of a nearby singularity. For example, suppose
\be
f(t)\Big|_{t\to t_0}  \sim  \phi(t)\, \Big(1-\frac{t}{t_0}\Big)^{-g}+\psi(t) \ ,  %\qquad t\to t_0 ,
\label{B.1}
\ee
where $\phi(t)$ and $\psi(t)$ are analytic near $t_0$. 
Then the Taylor expansion coefficients of $f(t)=\sum_k a_k t^k $ near the origin have large-order ($k\to\infty$) growth
\be
a_k\sim \frac{1}{t_0^k}
\binom{k+g-1}{k}
 \left[ \phi(t_0)- 
\frac{(g-1)\, t_0\, \phi^\prime(t_0)}{(k+g-1)}
+
\frac{(g-1)(g-2)\, t_0^2\, \phi^{\prime\prime}(t_0)}{2! (k+g-1)(k+g-2)}\, -\dots \right].
\label{B.2}
\ee
Thus,  leading and subleading  large-order behaviour terms determine the Taylor expansion of the analytic function $\phi(t)$ which multiplies the branch-cut factor in (\ref{B.1}). The function $\psi(t)$ can be extracted similarly by multiplying $f(t)$ through by $ \Big(1-\frac{t}{t_0}\Big)^{g}$, and applying the same procedure.
If the singularity is logarithmic,
\be
f(t)\Big|_{t\to t_0}  \sim  \phi(t)\, \ln\Big(1-\frac{t}{t_0}\Big)+\psi(t) \ ,  %\quad t\to t_0 ,
\label{B.3}
\ee
where $\phi(t)$ and $\psi(t)$ are analytic near $t_0$, 
then the Taylor expansion coefficients of $f(t)$ near the origin have large-order  ($k\to\infty$)  growth
\be
a_k\sim \frac{1}{t_0^k}\cdot \frac{1}{k} \left[\phi(t_0) - \frac{t_0\, \phi^\prime(t_0)}{(k-1)} +\frac{t_0^2\, \phi^{\prime\prime}(t_0)}{(k-1)(k-2)} -\dots \right]\ . 
\label{B.4}
\ee
This logarithmic behaviour  is found for the expansion coefficients of $\Delta F(\tilde\lambda)$, as shown in the right hand panel of Figure \ref{fig:trm1-ratio}.

\bibliography{BT-Biblio}
\bibliographystyle{JHEP}
\end{document}

 \iffa 
 &= \lim_{N\to \infty}N\,\bigg[2\pi^{2}\frac{\vev{\tr m^{2}e^{-S_{\rm int}}}-\vev{\tr m^{2}}\vev{e^{-S_{\rm int}}}}{\vev{e^{-S_{\rm int}}}}+\cdots\bigg]  \lp
 = -\lim_{N\to \infty} 2\pi^{2}N\bigg[\frac{\partial}{\partial x}\log\vev{e^{-S_{\rm int}}}+\cdots\bigg] = -\frac{\lambda^{2}}{4}\lim_{N\to\infty}\bigg[\frac{\partial}{\partial\lambda}\Delta F(\lambda; N) + \cdots\bigg]\lp
 = -\frac
 \be
\frac{\vev{\mc W}}{\vev{\mc W}^{\mc N=4}} = \frac{\int Dm\  \tr e^{2\pi m}e^{-S_{\rm int}}e^{-\frac{8\pi^{2}N}{\lambda}\tr m^{2}}}{\int Dm\   e^{-S_{\rm int}}e^{-\frac{8\pi^{2}N}{\lambda}\tr m^{2}}}
 \frac{\int Dm\  e^{-\tr m^{2}}}{\int Dm\   \tr e^{2\pi m}e^{-\frac{8\pi^{2}N}{\lambda}\tr m^{2}}} = \frac{\vev{\tr e^{2\pi m}e^{-S_{\rm int}}}}{\vev{\tr e^{2\pi m}}\, \vev{e^{-S_{\rm int}}}}
 \ee
 where $\vev{\cdots}$ in last equality are Gaussian averages with weight $e^{-\frac{8\pi^{2}N}{\lambda}\tr m^{2}}$. At large $N$ we have factorization and the ratio tends to 1.
  \be
\frac{\vev{\mc W}}{\vev{\mc W}^{\mc N=4}}  =  1+\frac{\vev{\tr e^{2\pi m}e^{-S_{\rm int}}}_{\rm conn}}{\vev{\tr e^{2\pi m}}\, \vev{e^{-S_{\rm int}}}}
 \ee
  \be
\frac{\vev{\mc W}}{\vev{\mc W}_{0}^{\mc N=4}}  =  \frac{\vev{\mc W}^{\mc N=4}}{\vev{\mc W}_{0}^{\mc N=4}}\bigg[1+\frac{\vev{\tr e^{2\pi m}e^{-S_{\rm int}}}_{\rm conn}}{\vev{\tr e^{2\pi m}}\, \vev{e^{-S_{\rm int}}}}\bigg]
 \ee
 Expanding in powers of $1/N$ gives 
 \be
 \Delta q = \lim_{N\to \infty}N\,\frac{\vev{\tr e^{2\pi m}e^{-S_{\rm int}}}_{\rm conn}}{\vev{\tr e^{2\pi m}}\, \vev{e^{-S_{\rm int}}}} = 
 \lim_{N\to \infty}N\,\bigg[\frac{\vev{\tr e^{2\pi m}e^{-S_{\rm int}}}}{\vev{\tr e^{2\pi m}}\, \vev{e^{-S_{\rm int}}}}-1\bigg].
 \ee
 Set $x = \frac{8\pi^{2}N}{\lambda}$. We have $\vev{e^{-S_{\rm int}}} = e^{-\Delta F(\lambda)}$. Formally expanding $\tr e^{2\pi m}$ the first non-vanishing term is 
 \ba
 \Delta q &= \lim_{N\to \infty}N\,\bigg[2\pi^{2}\frac{\vev{\tr m^{2}e^{-S_{\rm int}}}-\vev{\tr m^{2}}\vev{e^{-S_{\rm int}}}}{\vev{e^{-S_{\rm int}}}}+\cdots\bigg]  \lp
 = -\lim_{N\to \infty} 2\pi^{2}N\bigg[\frac{\partial}{\partial x}\log\vev{e^{-S_{\rm int}}}+\cdots\bigg] = -\frac{\lambda^{2}}{4}\lim_{N\to\infty}\bigg[\frac{\partial}{\partial\lambda}\Delta F(\lambda; N) + \cdots\bigg]\lp
 = -\frac{\lambda^{2}}{4}\frac{d}{d\lambda}\Delta F(\lambda),
 \ea
where we dropped dots from further expansion of $\tr e^{2\pi m}$. This is legitimate at large $N$ since these terms are expressed in terms of higher derivatives of $\Delta F(\lambda; N)$ 
with respect to $x\sim N$ and this  gives suppressed term.

\fi